\documentclass[floats,aps,showpacs,twocolumn,longbibliography,superscriptaddress]{revtex4-1}

\usepackage[dvips]{graphicx}
\usepackage{amsfonts}
\usepackage{amsmath,amssymb}
\usepackage{dsfont}

\usepackage{placeins}
\usepackage{afterpage}
\usepackage{flafter}
\usepackage[usenames,dvipsnames]{color}
\usepackage{lipsum}

\newcommand{\map}{U}
\newcommand{\qmap}{\widehat{\map}}

\newcommand{\Vkappa}{V_{\kappa}}
\newcommand{\Vepsilon}{V_{\epsilon}}

\newcommand{\erf}{\text{erf}}
\newcommand{\Arg}{\text{Arg}}
\newcommand{\myRe}{\text{Re}}
\newcommand{\myIm}{\text{Im}}

\newcommand{\width}{\xi}
\newcommand{\qabs}{q_{\text{abs}}}
\newcommand{\qmax}{q_{\text{max}}}
\newcommand{\pmax}{p_{\text{max}}}
\newcommand{\mmax}{m_{\text{max}}}
\newcommand{\xmax}{x_{\text{max}}}

\newcommand{\xb}{x_{\text{b}}}
\newcommand{\xf}{x_{\text{f}}}

\newcommand{\zf}{u_{0}}
\newcommand{\zfb}{\bar{u}_{0}}

\newcommand{\manifold}{W}
\newcommand{\sm}{\manifold^{s}}
\newcommand{\um}{\manifold^{u}}
\newcommand{\smb}{\bar{\manifold}^{s}}
\newcommand{\umb}{\bar{\manifold}^{u}}

\newcommand{\alphad}{\alpha}
\newcommand{\alphadbar}{\bar{\alpha}}
\newcommand{\alphadmax}{\alphad_{\max}}

\newcommand{\independent}{u}
\newcommand{\nui}{\beta_{\independent}}
\newcommand{\nuibar}{\bar{\beta}_{\independent}}
\newcommand{\nuimax}{\beta_{\independent}^{\max}}

\newcommand{\heff}{h_{\text{eff}}}

\newcommand{\qOp}{\hat{q}}
\newcommand{\pOp}{\hat{p}}
\newcommand{\HOp}{\widehat{H}}

\newcommand{\ket}[1]{\left| #1 \right\rangle}

\newcommand{\braOpket}[3]{\langle #1 | #2 | #3 \rangle}

\makeatletter
\let\Hy@backout\@gobble
\makeatother

\usepackage{color}

\begin{document}

\title{Resonance spectra for quantum maps of kicked scattering systems by
complex scaling}

\author{Normann Mertig}
\affiliation{Department of Physics, Tokyo Metropolitan University,
Minami-Osawa, Hachioji 192-0397, Japan}

\author{Akira Shudo}
\affiliation{Department of Physics, Tokyo Metropolitan University,
Minami-Osawa, Hachioji 192-0397, Japan}

\date{\today}

\begin{abstract}

We consider quantum maps induced by periodically-kicked scattering systems and
discuss the computation of their resonance spectra in terms of complex scaling
and sufficiently weak absorbing potentials.
We also show that strong absorptive and projective openings, as commonly used
for open quantum maps, fail to produce the resonance spectra of kicked
scattering systems, even if the opening does not affect the classical trapped
set.
The results are illustrated for a concrete model system whose dynamics resembles
key features of ionization and exhibits a trapped set which is organized by a
topological horseshoe at large kick strength.
Our findings should be useful for future tests of fractal Weyl conjectures and
investigations of dynamical tunneling.

\end{abstract}

\maketitle
\noindent

\section{Introduction}

Recent decades have witnessed significant progress in the semiclassical
understanding of resonance states and resonance spectra of complex scattering
systems.
In particular, for scattering systems with fully chaotic dynamics it was
explored
(i) how scattering from a chaotic repellor can lead to the formation of a
spectral gap \cite{GasRic1989a, GasRic1989b, GasRic1989c},
(ii) how properties of the classical dynamics are reflected in the fluctuations
of the scattering matrix \cite{BluSmi1989, BluSmi1990, Jun1990, JunPot1990,
JunTel1991},
(iii) how resonance eigenvalues can be determined from periodic orbits
\cite{CviEck1989, TanRicRos2000},
(iv) how counting functions of long-lived resonance states relate to the fractal
dimension of the classical invariant sets \cite{Sjo1990, Lin2002, LuSriZwo2003,
SchTwo2004, Non2011, Nov2013, KoeMicBaeKet2013}, and
(v) how resonance states localize on classical invariant sets
\cite{KeaNovPraSie2006, KoeBaeKet2015}.

In order to extend such results, for example to scattering systems with a mixed
phase space, it is crucial to have good toy models.
Ideally such toy models should be easy to exploit numerically and yet generic
enough to exhibit a robust effect of quantum-to-classical correspondence.
Toy models which exhibit these properties are \textit{open quantum maps}, see
Ref.~\cite{NonSjoZwo2011} for a definition, Refs.~\cite{Non2011, Nov2013} for
reviews, and Refs.~\cite{CasMasShe1997, CasGuaMas2000, SchTwo2004, NonZwo2005,
KeaNovPraSie2006, NonZwo2007, NonRub2007, KeaNovSch2008, KeaNonNovSie2008,
NovPedWisCarKea2009, KopSch2010, Non2011, IshAkaShuSch2012, LipRyuLeeKim2012,
Nov2013, KoeMicBaeKet2013, SchAlt2015, KoeBaeKet2015, MerKulLoeBaeKet2016,
FriBaeKetMer2017} for some examples.

In theory, open quantum maps quantize the the classical flow of an open system
along its trapped set and represent a reduction of a scattering system to a
sub-unitary operator, which holds the essential information on long-lived
resonances \cite{Non2011, NonSjoZwo2011, NonSjoZwo2011:p}.
In a similar spirit, one may think of open quantum maps as the time-evolution
operators corresponding to an effective Hamiltonian \cite{Sch2013b}, arising
from the Feshbach approach \cite{Fes1958, Fes1962, Rot2009} or a reduction of a
dielectric cavity \cite{KeaNovSch2008}.
Altogether, this shows that open quantum maps have a clear connection to
scattering systems and should therefore be generic.

In practice, however, none of the fore-mentioned approaches has lead to an open
quantum map which derives from a concrete scattering system.
Instead, there are many \textit{heuristic models} introduced in an ad-hoc
manner by combining the unitary time-evolution operator of a closed system,
such as a quantum map on a torus, with an absorption operator, such as a
Fresnel-type reflection operator or a projector.
See Refs.~\cite{CasMasShe1997, CasGuaMas2000, SchTwo2004, NonZwo2005,
KeaNovPraSie2006, NonZwo2007, NonRub2007, KeaNovSch2008, KeaNonNovSie2008,
NovPedWisCarKea2009, KopSch2010, Non2011, IshAkaShuSch2012, LipRyuLeeKim2012,
Nov2013, KoeMicBaeKet2013, SchAlt2015, KoeBaeKet2015, MerKulLoeBaeKet2016,
FriBaeKetMer2017} for examples.
While these heuristic models have stimulated many findings in the recent past,
they do not represent a step by step reduction of a scattering system in the
sense of Ref.~\cite{NonSjoZwo2011} and it remains unclear to which degree they
faithfully represent resonances.

Another somewhat distracting aspect of heuristic models is the use of
non-analytic absorption operators, because:\
(i) The reduction of scattering systems to open quantum maps as described in
\cite{Non2011}, explicitly excludes non-analytic absorption operators.\
(ii) In quantum chemistry computations of resonance spectra, based on absorbing
potentials, one carefully has to minimize the back-reflections from absorbing
potentials into the scattering region \cite{RisMey1993}.
In such cases rapidly varying absorbing potentials with non-analytic parts
obscure the computation.\
(iii) Scattering systems with non-analytic potentials usually exhibit
diffraction effects which makes quantitative investigations of
quantum-to-classical correspondence considerably more complicated
\cite{VatWirRos1994, BreWirRotStaBur2010, LacBreBurLib2013}.
This makes quantum map models of scattering based on analytic potentials
desirable.
\begin{figure*}[!tb]
  \begin{center}
    \includegraphics[]{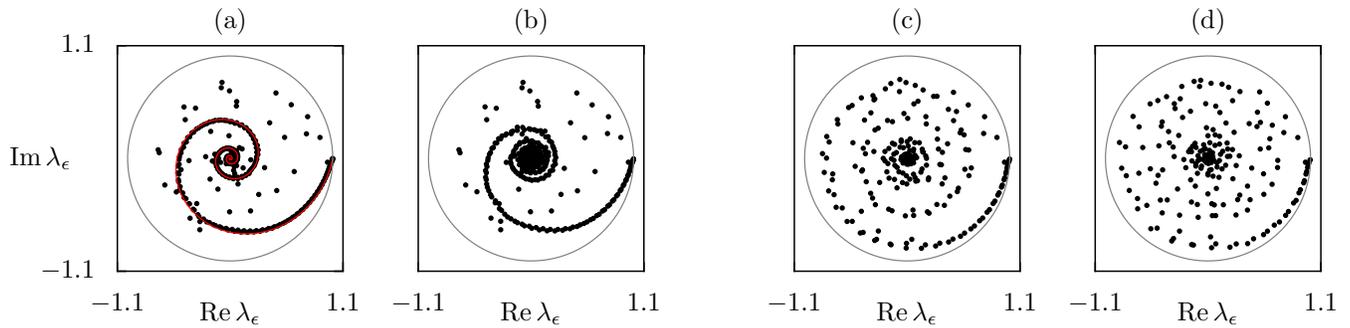}
     \caption{(color online) Numerically determined spectra (dots) of a
kicked scattering system for kick strength $\kappa=11$ (chaotic dynamics) and
effective Planck constant $\heff=1/50$.
See text for details.
The gray line represents a unit circle.
(a,b) For (a) complex scaling and (b) weak absorbing potentials the resonance
spectrum separates from the continuous spectrum.
The continuous spectrum localizes along a spiral ([red] line in (a)).
(c,d) Numerical spectra obtained with (c) strong absorbing potentials and (d)
projective openings.
The separation of continuous and resonance spectra is lost.
}
     \label{fig:spectrum_box_VABS_CS}
  \end{center}
\end{figure*}

In this paper, we consider quantum maps induced by periodically-kicked
scattering systems with analytic potentials.
In contrast to heuristic models this naturally accommodates resonances and does
not require adding any absorption.
In order to compute the resonance spectra of such quantum maps, we adapt the
method of complex scaling \cite{AguCom1971, BalCom1971, ReeSim1978, Rei1982b,
Moi1998} as previously developed for time-periodic systems \cite{ChuRei1977,
MaqChuRei1983, MoiKor1990, BenMoiLefKos1991, BenMoiKosCer1993}.
This gives the complex-scaled time-evolution operator,
Eq.~\eqref{eq:PropagatorScaled}, i.e., the \textit{complex-scaled
quantum map} from which the resonance spectrum can readily be computed, see
Fig.~\ref{fig:spectrum_box_VABS_CS}(a) for an example.

We further examine the connection of the complex-scaled time-evolution operator
with heuristic models.
To this end we adapt the method of absorbing potentials \cite{RisMey1993} to
kicked scattering systems.
This results in an absorption-augmented time-evolution operator,
Eq.~\eqref{eq:PropagatorAbsorption}, which has the same structure as heuristic
model systems.
We show that this operator also allows for computing the resonance spectrum,
albeit only in the limit of sufficiently weak absorption, see
Fig.~\ref{fig:spectrum_box_VABS_CS}(b).
On the other hand, we demonstrate that strong absorbing potentials and
projective openings, as commonly used for heuristic models, do not allow for
computing the resonance spectrum of kicked scattering systems, see
Fig.~\ref{fig:spectrum_box_VABS_CS}(c,d), respectively.

In this context a novel point, which was previously ignored in heuristic
models, is the appearance of a continuous spectrum, which must be carefully
separated from the resonance spectrum in numerical computations \cite{Moi1998,
RisMey1993}.
This separation can be accomplished by means of complex scaling and weak
absorbing potentials.
On the other hand, both for strong absorption and projective openings, we
observe a blow up of the continuous spectrum which obscures its effective
separation from resonance eigenvalues, see Fig.~\ref{fig:spectrum_box_VABS_CS}.
We emphasize that this effect occurs even if the absorption does not affect the
trapped set, which carries the semiclassical information on the resonance
spectrum.

In order to illustrate our results with numerically computed spectra, we
introduce a concrete \textit{model system}, based on kicking potentials which
are entire functions.
For this model system we also discuss the organization of its classical flow in
terms of few stable and unstable manifolds, following standard methods in
Hamiltonian transport \cite{MacMeiPer1984, MacMeiPer1984b, Rom1990, Wig1992,
Rom1994}.
The purpose of this discussion is to emphasize the structural similarity of the
model system with phase-space structures which commonly appear in classical
ionization \cite{MitHanTigFloDel2004, MitHanTigFloDel2004b, BurMitWykYeDun2011}
and dissociation \cite{DavGra1986}.
Moreover, we use these structures to show that the trapped set of the model
system is organized by a topological horseshoe at large kick strength.

Ultimately, we hope that the uniform hyperbolicity of the proposed model system
can be proven.
This would result in a new sibling among the uniformly hyperbolic model
systems, which, in contrast to the three-disc scatterer \cite{Eck1987,
GasRic1989a, GasRic1989b, GasRic1989c} and the open baker map \cite{NonZwo2005,
NonZwo2007, NonRub2007, KeaNonNovSie2008, Non2011}, is free from distracting
diffraction effects.
This should be useful for future investigations of fractal Weyl conjectures and
further topics in chaotic scattering.
Even more so, we expect that the proposed model system should be useful for
exploiting scattering systems with a mixed phase space, for which many more
questions of quantum-to-classical correspondence remain unsolved.

The body of this paper is preceded by Sec.~\ref{Sec:KickedScatteringSystem},
in which we specify periodically-kicked scattering systems and lay out some
basic notation.
In Sec.~\ref{Sec:ModelSystemClassical} we introduce the model system and
discuss its classical dynamics.
In Sec.~\ref{Sec:ComputingResonances} we derive quantum maps for
periodically-kicked scattering systems and discuss the computation of their
resonance spectra in terms of complex scaling and absorbing potentials.
The summary and discussion are given in Sec.~\ref{Sec:SummaryAndDiscussion}.
The numerical computations are documented in the Appendix.

\section{Kicked scattering systems}
\label{Sec:KickedScatteringSystem}

In this paper we are concerned with periodically-kicked, one-degree of freedom
quantum systems which obey the time-dependent Schr\"odinger equation
\begin{align}
  \label{eq:SchroedingerEquation}
\left[
 i\hbar\frac{\partial}{\partial t}
 +\frac{\hbar^2}{2}\frac{\partial^2}{\partial q^{2}}
 - V(q)\sum_{n\in\mathbb{Z}}\delta(t-n)
\right] \psi(q,t) =0.
\end{align}
Here, all quantities have been scaled to dimensionless units.
In particular, $q\in\mathbb{R}$ denotes the position, $t\in\mathbb{R}$ denotes
time, $\hbar\in\mathbb{R}_{+}$ denotes the reduced effective Planck constant,
and $\psi$ represents the wave function.
We further refer to $\heff:=2\pi\hbar$ as the effective Planck constant.
It is treated as a free parameter and $\heff\to0$ denotes the semiclassical
limit.
Furthermore, $\delta(\cdot)$ represents the Dirac $\delta$ function.

The corresponding classical dynamics is given by the Hamilton function
\begin{align}
  \label{eq:HamiltonFunction}
  H(q,p;t) = \frac{p^2}{2} + V(q)\sum_{n\in\mathbb{Z}}\delta(t-n),
\end{align}
with $p\in\mathbb{R}$ being the momentum.

The crucial point of this paper is to consider kicking potentials $V(q)$ which
tend to zero sufficiently fast as $|q|$ tends to infinity
\begin{align}
  \label{eq:DecayingPotential}
  \lim_{q \to \pm\infty} V(q) = 0,
\end{align}
i.e., we consider scattering systems for which the classical dynamics resemble
free motion as $|q|$ tends to infinity.

\textit{Classical map} -- In order to investigate the classical dynamics of
periodically-kicked scattering systems we use stroboscopic phase-space sections.
They are obtained by integrating Hamilton's equations of motion over one period
in time.
This results in a symplectic map on $\mathbb{R}^{2}$
\begin{align}
  \label{eq:StroboscopicMap}
    \map: (q_n, p_n) \mapsto (q_{n+1}, p_{n+1}),
\end{align}
which in its symmetrized version is given by
\begin{subequations}
  \label{eq:HalfKick}
  \begin{align}
    q_{n+1} =&\, q_n + p_n - \frac{1}{2} V'(q_n), \label{eq:HalfKickPos} \\
    p_{n+1} =&\, p_n  - \frac{1}{2} V'(q_n) - \frac{1}{2}V'(q_{n+1}).
\label{eq:HalfKickMom}
  \end{align}
\end{subequations}
Here, $(q_n, p_n)$ denotes phase-space points along a trajectory in the middle
of the $n$th kick, while $V'$ denotes the derivative of the kicking potential
with respect to $q$.

\textit{Quantum map} -- The quantum mechanical analogue of the classical map
$\map$ is the time-evolution operator over one period of the external driving
\cite{BerBalTabVor1979, CasChiIzrFor1979}.
This operator is referred to as quantum map \cite{BerBalTabVor1979}.
It is given by
\begin{align}
  \label{eq:Qmap}
 \hspace*{-0.2cm} \qmap =
    \exp\left(\!-\frac{i}{\hbar} \frac{V(\qOp)}{2}\right)
    \exp\left(\!-\frac{i}{\hbar} \frac{\pOp^2}{2} \right)
    \exp\left(\!-\frac{i}{\hbar} \frac{V(\qOp)}{2}\right),
\end{align}
where the operators fulfill the usual commutation relation $[\qOp, \pOp] =
i\hbar$.
Due to Eq.~\eqref{eq:DecayingPotential} the system is unbound and the quantum
map supports a resonance spectrum.
In that, it is already \textit{open}.
In order to access its resonance spectrum we will derive its complex-scaled
form, Eq.~\eqref{eq:PropagatorScaled}, and its absorption-augmented form,
Eq.~\eqref{eq:PropagatorAbsorption}.

\section{Model system: Classical dynamics}
\label{Sec:ModelSystemClassical}

In this section we specify the model system and discuss its classical dynamics.

\subsection{A family of kicking potentials}

In what follows, we consider maps induced by the kicking potential
\begin{align}
  \label{eq:KickingPotential}
  V(q) = \Vkappa(q) + \Vepsilon(q).
\end{align}
Its main part is an attractive Gaussian
\begin{align}
  \label{eq:GaussianKick}
  \Vkappa(q) = -\frac{\kappa}{16}\exp{(-8 q^{2})},
\end{align}
with kick strength $\kappa\geq 0$.
This system has previously been studied as a model of chaotic scattering
\cite{Jen1992, Jen1995} as well as a model of a pulsed optical trap for cold
atom systems \cite{KriFisOttAnt2011}.
For $\kappa<0$ it has further been used as a model of an oscillating barrier
potential \cite{OniShuIkeTak2001, OniShuIkeTak2003}.
The dimensionless units are chosen such that a version of Chirikov's standard
map \cite{Chi1979}, induced by $V(q)=-\kappa/(2 \pi)^{2}\cos(2\pi q)$ is
mimicked in the region $q\in(-0.5,0.5)$.
This applies in particular to the stability matrix which controls the dynamics
around the central fixed point $(q_{\text{c}}, p_{\text{c}}) = (0, 0)$ as well
as the extrema of $V'(q)$, located at $q_{\text{e}}=\pm1/4$.

In order to confine the trapped set in a compact region we propose the
perturbation term
\begin{align}
  \label{eq:PerturbationKick}
  \Vepsilon(q) = -\epsilon
              \left[ \erf{\left(\sqrt{8}\left[q-\xb\right] \right)} -
              \erf{\left(\sqrt{8}\left[q+\xb\right] \right)} \right],
\end{align}
with $\erf(\cdot)$ being the error function, $\epsilon\geq0$ a small
perturbation strength and $\xb$ a positive real parameter.
Its purpose is to increase the energy for $q\in(-\xb,\xb)$ which ensures that
trajectories which escape beyond a certain point do not return.
In particular, we use the perturbation strength
\begin{align}
  \label{eq:PerturbationStrength}
  \epsilon = \frac{\kappa \sqrt{\pi}}{2\sqrt{8}}
\,\xf\,\frac{\exp\left(-8\xb\left[2\xf-\xb\right]\right)}{1
- \exp\left(-32\xf\xb\right)}.
\end{align}
with
\begin{align}
  \label{eq:ParameterConditions}
  16\xb\xf > 1.
\end{align}

If condition~\eqref{eq:ParameterConditions} is satisfied any trajectory entering
the outgoing regions
\begin{align}
  \label{eq:OutgoingRegionPlus}
  \mathcal{O}^{+} &= \{(q,p)\in\mathbb{R}^{2}\quad\text{:}\quad p>0, q>\xf\} \\
  \label{eq:OutgoingRegionMinus}
  \mathcal{O}^{-} &= \{(q,p)\in\mathbb{R}^{2}\quad\text{:}\quad p<0, q<-\xf\}
\end{align}
escapes to $q\to\pm\infty$, respectively.
Due to time reversal symmetry, this further gives the incoming regions
\begin{align}
  \label{eq:IncomingRegionPlus}
  \mathcal{I}^{+} &= \{(q,p)\in\mathbb{R}^{2}\quad\text{:}\quad p<0, q>\xf\} \\
  \label{eq:IncomingRegionMinus}
  \mathcal{I}^{-} &= \{(q,p)\in\mathbb{R}^{2}\quad\text{:}\quad p>0, q<-\xf\}
\end{align}
where trajectories escape to $q\to\pm\infty$ in backward time.
See Fig.~\ref{fig:manifold_K29} for a sketch and Appendix~\ref{App:EscapeRegion}
for details.
In that, condition~\eqref{eq:ParameterConditions} also ensures that the
trapped set
\begin{align}
 \label{eq:TrappedSet}
 \mathcal{K}:=\{(q,p)\in\mathbb{R}^2\quad:\quad\lim_{n\to\pm\infty}
|U^{n}(q,p)|<\infty \},
\end{align}
is confined to the strip $(q,p)\in[-\xf,\xf]\times\mathbb{R}$.
Even more, one can easily show that the trapped set must be contained in a
compact square
\begin{align}
 \mathcal{K} \subset [-\xf,\xf]\times[-\pmax,\pmax].
\end{align}
Here, $\pmax$ must be chosen sufficiently large, such that any point with
$q\in[-\xf,\xf]$ and $|p|>\pmax$ has sufficient momentum to reach either
of the outgoing regions $\mathcal{O}^{\pm}$ in one iteration of the map $\map$.
Since $V'(q)$ is bounded, a valid choice is given by
$\pmax=2\xf+\max_{q\in\mathbb{R}}\{V'(q)\}$.

The above choice of parameters, Eqs.~\eqref{eq:PerturbationStrength}
and \eqref{eq:ParameterConditions}, further allows for locating two unstable
fixed-points at\
\begin{align}
\label{eq:FixedPoints}
 \zf := (\xf, 0) \quad \text{and} \quad \zfb := (-\xf, 0).
\end{align}
As discussed in the next section their stable and unstable manifolds control the
dynamics in the scattering region and provide sharper bounds on the trapped set.

Note that the proposed kicking potential fulfills
Eq.~\eqref{eq:DecayingPotential}.
In fact, it decreases to zero exponentially fast and drops below machine
precision for $|q|\gtrsim2$.
It further is an entire function, i.e., it has no discontinuities.

\subsection{Phase space}

From now on we fix the parameters of the model system as $\xf=1.2$ and $\xb=1$
and discuss its phase space.

\subsubsection{Scattering region}
\label{Sec:ScatteringRegion}

\begin{figure}[tb]
  \begin{center}
    \includegraphics[]{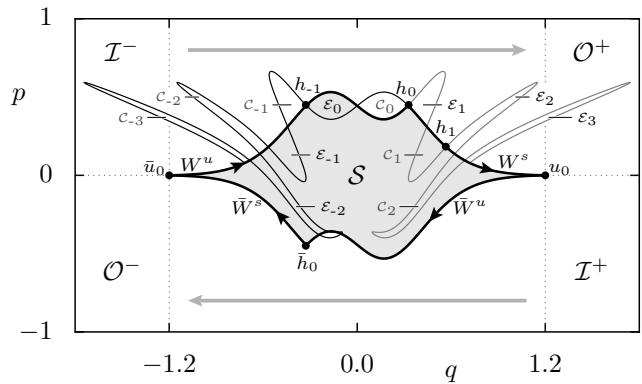}
     \caption{Phase space of the model system for $\kappa=2.9$.
The initial segments of the stable and unstable manifolds (thick lines) form the
boundary of the (gray shaded) scattering region.
Dots represent fixed points and heteroclinic points.
Thin lines show the stable (black) and unstable (gray) manifolds, which form
the boundary of the lobe areas.
Gray arrows indicate transport from incoming to outgoing regions, as marked by
dotted lines.
}
     \label{fig:manifold_K29}
  \end{center}
\end{figure}

In this section we discuss how the stable and unstable manifolds of the fixed
points, $\zf$ and $\zfb$, organize the classical dynamics.
The discussion follows standard arguments in classical transport
\cite{MacMeiPer1984, MacMeiPer1984b, Rom1990, Wig1992, Rom1994} and commonly
appears in problems of classical ionization \cite{MitHanTigFloDel2004,
MitHanTigFloDel2004b, BurMitWykYeDun2011} and dissociation \cite{DavGra1986}.
The discussion is visually supported by Fig.~\ref{fig:manifold_K29} for the
example $\kappa=2.9$.
Qualitatively similar structures are observed for any $\kappa>0$.

In what follows, we omit manifolds which trivially extend into the outgoing and
incoming regions.
We further focus on one side of the symmetry.
Related properties for the symmetry partners, as marked by an overbar, are
implied.
We refer to the stable manifold of $\zf$ as $\sm$ and the unstable manifold of
$\zfb$ as $\um$.
A closed segment of a manifold $\manifold$ with endpoints $x,x'$ is
referred to as $\manifold[x,x']$.

\textit{Initial segments} --
We start by numerically constructing the initial segments of the relevant
manifolds up to their first heteroclinic intersection points, $h_{0}$ and
$\bar{h}_{0}$.
Here, $h_{0}$ is chosen such that
\begin{align}
 \label{eq:HeteroclinicIntersection}
 \um[\zfb, h_{0}] \cap \sm[h_{0}, \zf]= h_{0}.
\end{align}
Note that $h_{0}$ gives rise to a heteroclinic orbit and replacing it by any of
its forward or backward iterates is a topologically equivalent choice.
The example system exhibits two distinct heteroclinic orbits, whose points
satisfy condition~\eqref{eq:HeteroclinicIntersection}.
Here, we choose $h_{0}$ from the heteroclinic orbit which has no point along the
symmetry axis $q=0$.
This choice is convenient in the discussion of the topological horseshoe further
below.

\textit{Scattering region} --
We now define the scattering region $\mathcal{S}$.
It is the compact set, whose boundary is given by the closed curve of initial
segments
\begin{align}
 \partial\mathcal{S} = \um[\zfb,h_{0}] \cup \sm[h_{0}, \zf] \cup \umb[\zf,
\bar{h}_{0}] \cup \smb[\bar{h}_{0},\zfb], \nonumber
\end{align}
see the thick lines and the gray shaded region in Fig.~\ref{fig:manifold_K29}.

\textit{Lobes} -- In the next step we consider a backward iteration of the map
$\map$.
This maps the heteroclinic point $h_{0}$ onto $h_{-1}$.
It further shortens the initial segment of the unstable manifold to the point
$h_{-1}$.
The initial segment of the stable manifold reproduces up to the point $h_{0}$
and produces an additional segment $\sm[h_{0},h_{-1}]$.
This segment together with the segment $\um[h_{0},h_{-1}]$ form the boundary of
two lobes, labeled as $\mathcal{E}_{0}$ and $\mathcal{C}_{0}$ in
Fig.~\ref{fig:manifold_K29}.
To be precise, $\mathcal{E}_{0}$ shall include boundary points which are not
on the stable segment, while $\mathcal{C}_{0}$ shall include boundary points
which are not on the unstable segment.
Propagating the lobes and the corresponding segments of the stable and unstable
manifold forward and backward to times $i\in\mathbb{Z}$ gives the lobe
iterates, according to
\begin{subequations}
 \label{eq:LobePropagation}
 \begin{align}
 \mathcal{C}_{i} &= \{ (q,p)\in\mathbb{R}^2 \quad : \quad
\map^{-i}(q,p)\in\mathcal{C}_{0}\}, \\
 \mathcal{E}_{i} &= \{ (q,p)\in\mathbb{R}^2 \quad : \quad
\map^{-i}(q,p)\in\mathcal{E}_{0}\}.
 \end{align}
\end{subequations}
This is depicted in Fig.~\ref{fig:manifold_K29}.

\textit{Partial barrier, flux, and turnstiles} -- As discussed in
Refs.~\cite{MacMeiPer1984, MacMeiPer1984b, Wig1992}, the curve $\um[\zfb,
h_{0}] \cap \sm[h_{0}, \zf]$ is a partial barrier.
Orbits which leave the scattering region across the partial barrier in one
iteration of the map are initially in $\mathcal{E}_{0}$ and map to
$\mathcal{E}_{1}$.
Orbits which enter the scattering region across the partial barrier in one
iteration of the map are initially in $\mathcal{C}_{0}$ and map to
$\mathcal{C}_{1}$.
The Lebesgue measure $\mu(\cdot)$ of all lobes is equal
\begin{align}
 \Phi:= \mu(\mathcal{E}_{i}) = \mu(\mathcal{C}_{i}) \quad \forall i \in
\mathbb{Z}, \nonumber
\end{align}
such that $\Phi$ denotes the flux exchanged across the partial barrier in one
iteration of the map.
This mechanism is called turnstile transport.

\textit{No return, escape, and capture} -- The model system exhibits
additional properties, which commonly appear in classical ionization
\cite{MitHanTigFloDel2004, MitHanTigFloDel2004b} and dissociation
\cite{DavGra1986}.
Namely, after leaving the scattering region the $\mathcal{E}$-lobes do not
return and propagate into the outgoing region
\begin{align}
 \label{eq:escapeforward}
 \mathcal{S}\cap\mathcal{E}_{i>0} = \emptyset \quad \text{and}
\quad \lim_{i\to\infty} \mathcal{E}_{i} \subset \mathcal{O}^{+}.
\end{align}
Hence, we call them escape lobes.
Similarly, after leaving the scattering region in backward time the
$\mathcal{C}$-lobes do not return and propagate into the incoming region
\begin{align}
 \label{eq:escapebackward}
 \mathcal{S}\cap\mathcal{C}_{i\leq0} = \emptyset \quad \text{and}
\quad \lim_{i\to-\infty} \mathcal{C}_{i} \subset \mathcal{I}^{-}.
\end{align}
Since any orbit propagating from an incoming region into the scattering region
must pass through $\mathcal{C}_{0}$, we call them capture lobes.

\textit{Trapped set} -- The above properties imply that the trapped set
$\mathcal{K}$ is confined to the scattering region as
\begin{align}
 \mathcal{K} \subset \mathcal{S} \setminus \cup _{i\in\mathbb{Z}}
\left(\mathcal{C}_{i} \cup \mathcal{E}_{i} \cup \bar{\mathcal{C}}_{i} \cup
\bar{\mathcal{E}}_{i} \right).
\end{align}
This is because, any orbit of the model system which is neither on a capture
lobe nor within the scattering region, trivially propagates towards the outgoing
region.
Hence, the trapped set must be on the capture lobes or within the scattering
region.
Further, excluding both escape and capture lobes, due to
Eq.~\eqref{eq:escapeforward} and Eq.~\eqref{eq:escapebackward}, gives the above
property.

\textit{Dynamics in the scattering region} --
For intermediate kick strength ($1\lesssim\kappa\lesssim6$) the dynamics in the
scattering region exhibit a mixed phase space.
Here, the trapped set $\mathcal{K}$ accommodates regular motion along invariant
tori.
As the kicking strength of the model system is increased beyond $\kappa\gtrsim
8$, we are no longer able to detect regular motion in numerical simulations.
In particular, when searching for periodic orbits up to period ten, we find
all of the numerically detected orbits to be unstable.
The dynamics in the scattering region becomes chaotic.

\subsubsection{Topological horseshoe}
\label{Sec:Chaos}
\begin{figure}[tb]
  \begin{center}
    \includegraphics[]{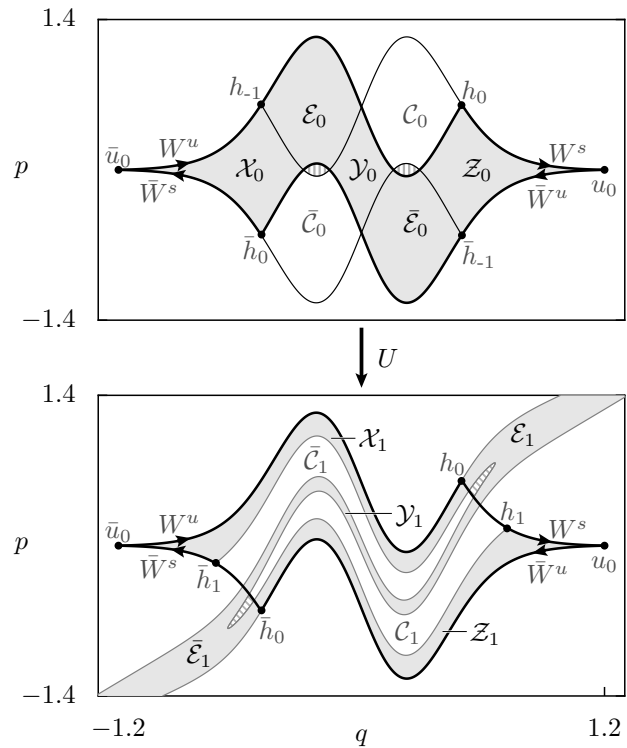}
     \caption{The scattering region and its forward iterate (gray areas), for
$\kappa=11$.
Lines represent the stable and unstable manifolds of the fixed points $\zf$,
$\zfb$.
Dots denote fixed points and heteroclinic points.
Gray striped regions indicate the non-empty intersections,
Eq.~\eqref{eq:HorseshoeConditionPropagated}.
}
     \label{fig:manifold_K11}
  \end{center}
\end{figure}
\begin{figure}[tb]
  \begin{center}
    \includegraphics[]{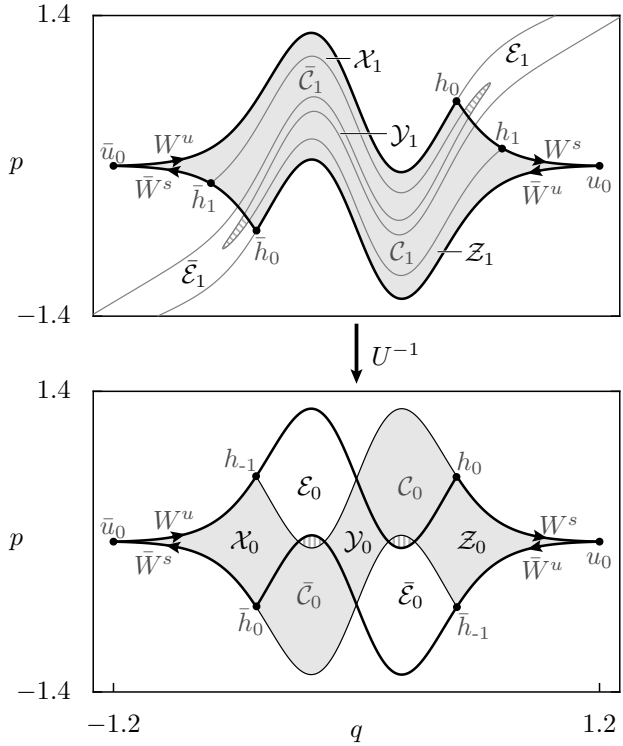}
     \caption{The scattering region and its backward iterate (gray areas), for
$\kappa=11$.
Lines represent the stable and unstable manifolds of the fixed points $\zf$,
$\zfb$.
Dots denote fixed points and heteroclinic points.
Gray striped regions indicate the non-empty intersections,
Eq.~\eqref{eq:HorseshoeConditionPropagated}.
}
     \label{fig:manifold_K11_backward}
  \end{center}
\end{figure}

In this section we present numerical evidence that the scattering region turns
into a topological horseshoe for kick strength $\kappa\gtrsim 10.5$.

We start by considering the forward and backward iterates of the scattering
region $\mathcal{S}$, as is illustrated in Figs.~\ref{fig:manifold_K11} and
\ref{fig:manifold_K11_backward} for $\kappa=11$.
Here, the forward iterate of the scattering region $\map(\mathcal{S})$ cuts upon
itself along three mutually disjoint sets $\mathcal{X}_{1}, \mathcal{Y}_{1},
\mathcal{Z}_{1}$
\begin{align}
 \mathcal{S}\cap\map(\mathcal{S}) = \mathcal{X}_{1} \cup \mathcal{Y}_{1} \cup
\mathcal{Z}_{1}.
\end{align}
The rest of the forward iterate $\map(\mathcal{S})$ locates along the escape
lobes, $\mathcal{E}_{1}$ and $\bar{\mathcal{E}}_{1}$, which do not return to the
scattering region, see Fig.~\ref{fig:manifold_K11}.
Similarly, the backward iterate of the scattering region
$\map^{-1}(\mathcal{S})$ cuts upon itself along three mutually disjoint sets
$\mathcal{X}_{0}, \mathcal{Y}_{0}, \mathcal{Z}_{0}$ %
\begin{align}
 \mathcal{S}\cap\map^{-1}(\mathcal{S}) = \mathcal{X}_{0} \cup \mathcal{Y}_{0}
\cup \mathcal{Z}_{0}.
\end{align}
The rest of the backward iterate locates along the capture lobes,
$\mathcal{C}_{0}$ and $\bar{\mathcal{C}}_{0}$, which do not return
to the scattering region in backward time, see
Fig.~\ref{fig:manifold_K11_backward}.
This is a topological horseshoe.

\begin{figure}[tb]
  \begin{center}
    \includegraphics[]{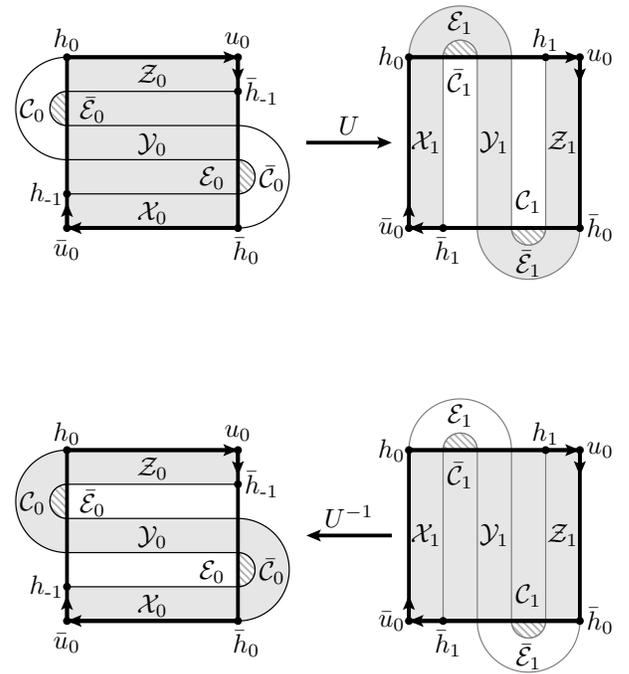}
     \caption{The scattering region and its forward (upper) and backward
(lower) iterate are schematically shown by gray areas.
Lines represent the stable and unstable manifolds of the fixed points $\zf$,
$\zfb$.
Dots represent fixed points and heteroclinic points.
Gray striped regions indicate the non-empty intersections,
Eq.~\eqref{eq:HorseshoeConditionPropagated}.
}
     \label{fig:mapping_K11_schematic}
  \end{center}
\end{figure}
In order to show the structure of the topological horseshoe more clearly, we
summarize the foregoing discussion in a homeomorphically equivalent, schematic
representation in Fig.~\ref{fig:mapping_K11_schematic}.
Here, the scattering region appears as a square in a plane.
Furthermore, any point which leaves the square, either in forward or backward
time, tends to infinity and never returns.
The forward and backward iterates of the scattering region cut across itself in
three mutually disjoint vertical and horizontal stripes, thus forming a
topological horseshoe.

\begin{figure}[tb]
  \begin{center}
    \includegraphics[]{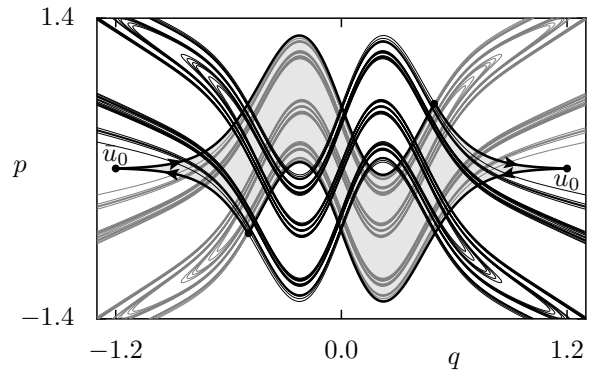}
     \caption{Stable and unstable manifolds of the fixed points $\zf$,
$\zfb$ for $\kappa=11$ are depicted by gray and black lines, respectively.
The initial segments (thick lines) form the boundaries of the (gray shaded)
scattering region.
}
     \label{fig:manifold_K11_uh}
  \end{center}
\end{figure}
We remark that the occurrence of the topological horseshoe is related with a
phenomenon reminiscent of the first tangency in the H{\'e}non map.
Namely, for the model system, we found that the scattering region turns into a
topological horseshoe, if the following lobes have non-empty intersections
\begin{align}
 \label{eq:HorseshoeCondition}
 \mathcal{E}_{0}\cap\bar{\mathcal{C}}_{0} \neq
\emptyset \quad\text{ and } \quad
 \bar{\mathcal{E}}_{0}\cap\mathcal{C}_{0} \neq \emptyset.
\end{align}
More specifically, if the corresponding boundary segments of the stable and
unstable manifold intersect transversally in exactly two homoclinic points.
See gray striped regions in Figs.~\ref{fig:manifold_K11},
\ref{fig:manifold_K11_backward} and \ref{fig:mapping_K11_schematic}.
This intersection enforces a topological horseshoe because, combining
Eq.~\eqref{eq:HorseshoeCondition} with Eq.~\eqref{eq:LobePropagation}
immediately implies that
\begin{align}
 \label{eq:HorseshoeConditionPropagated}
 \forall i\in\mathbb{Z}: \quad \mathcal{E}_{i}\cap\bar{\mathcal{C}}_{i} \neq
\emptyset \quad\text{ and } \quad
 \bar{\mathcal{E}}_{i}\cap\mathcal{C}_{i} \neq \emptyset.
\end{align}
This intersection of escape and capture lobes implies that the capture lobes
$\bar{\mathcal{C}}_{0}$ and $\mathcal{C}_{0}$ are pulled through the scattering
region in a forward iteration.
As illustrated in Figs.~\ref{fig:manifold_K11} and
\ref{fig:mapping_K11_schematic}(upper part), this results in capture lobes
$\bar{\mathcal{C}}_{1}$ and $\mathcal{C}_{1}$ which cut the remaining part of
the scattering region into three disjoint stripes.
Similarly, the non-empty intersections of
Eq.~\eqref{eq:HorseshoeConditionPropagated} for $i=1$, imply that the escape
lobes $\bar{\mathcal{E}}_{1}$ and $\mathcal{E}_{1}$ are pulled through the
scattering region in a backward iteration.
As illustrated in Figs.~\ref{fig:manifold_K11_backward} and
\ref{fig:mapping_K11_schematic}(lower part), this results in escape lobes
$\bar{\mathcal{E}}_{0}$ and $\mathcal{E}_{0}$ which cut the remaining part of
the scattering region into three disjoint stripes.
This enforces a topological horseshoe.
Condition~\eqref{eq:HorseshoeCondition} can be checked numerically, using finite
segments of the stable and unstable manifolds.
We found that it is fulfilled for any tested value $\kappa>10.5$.

\textit{Complete horseshoe and uniform hyperbolicity?} --
Note that a topological horseshoe is not yet a sufficient condition for a
complete horseshoe \cite{Sma1962, Wig2003}.
Proving a complete horseshoe might be achieved by checking the sector condition,
described in Ref.~\cite{Wig2003}, but was not attempted here.
Note that a topological horseshoe further cannot guarantee the uniform
hyperbolicity of the model system.
Numerical evidence in favor of uniform hyperbolicity is obtained by constructing
long but finite approximations to the stable and unstable manifolds.
For any tested value $\kappa>10.5$ we observe that these manifolds always
intersect at finite angles.
See Fig.~\ref{fig:manifold_K11_uh} for the case $\kappa=11$.

\section{Resonance spectra and quantum maps}
\label{Sec:ComputingResonances}

In this section we derive quantum maps for kicked scattering systems and discuss
the computation of their resonance spectra in terms of complex scaling and
absorbing potentials.

In particular, in Sec.~\ref{Sec:ComplexScaling} the method of complex scaling
\cite{AguCom1971, BalCom1971, ReeSim1978, Rei1982b, Moi1998} as developed for
time-periodic systems \cite{ChuRei1977, MaqChuRei1983, MoiKor1990,
BenMoiLefKos1991, BenMoiKosCer1993} is adapted to periodically-kicked scattering
systems.
This results in the complex-scaled time-evolution operator,
Eq.~\eqref{eq:PropagatorScaled}, i.e., the complex-scaled quantum map from
which resonance spectra can readily be computed, see Fig.~\ref{fig:spectrum_CS}.
In Sec.~\ref{Sec:AbsorbingPotential}, we further examine resonance spectra of
heuristic model systems.
To this end we adapt the method of complex absorbing potentials
\cite{RisMey1993} to kicked scattering systems.
This leads to an absorption-augmented time-evolution operator,
Eq.~\eqref{eq:PropagatorAbsorption}, which admits the form of a heuristic model.
We show that this operator also allows for computing the resonance spectra, see
Fig.~\ref{fig:spectrum_VABS}(b), albeit only in the limit of weak absorption.
On the other hand, we also show that strong absorption,
Fig.~\ref{fig:spectrum_VABS}(a), and projective openings,
Fig.~\ref{fig:SpectrumBox}, as commonly applied to heuristic models, fail to
produce the resonance spectrum.
An overview of the main ideas is given in Sec.~\ref{Sec:Overview}, while
introductory remarks on Floquet theory and complex scaling are given in
Sec.~\ref{Sec:Floquet} and Sec.\ref{Sec:ComplexScalingTimeIndependent},
respectively.

\subsection{Computation of resonances: Overview}
\label{Sec:Overview}

The main goal of Sec.~\ref{Sec:ComputingResonances} is to numerically compute
resonance spectra of periodically-kicked scattering systems,
Eq.~\eqref{eq:SchroedingerEquation}.
Since this task can at times become quite technical, we give an outline of the
key points and recommend Ref.~\cite{Moi1998} for a review.

\textit{Resonances} -- Since periodically-kicked scattering systems are unbound,
they admit resonance solutions with eigenvalues in the lower half of the complex
plane.
Their real parts denote (quasi-)energies, while their imaginary part describes
the life-times of a resonance solution and may, for example be associated with
the width of a spectral line.
Mathematically, a resonance is associated with the poles of the resolvent,
analytically continued into the lower half of the complex plane.
In that, resonance states are to a scattering system as bound states to a closed
system, i.e., they allow for efficient basis expansions of wave-packets,
scattering cross-sections, or the scattering matrix.
See Ref.~\cite{Moi1998} for an overview.

\textit{Challenges} -- In numerical computations of resonance spectra, we face
several challenges which are unfamiliar from closed systems:\
(i) Resonance states or not square-integrable. Even worse, they diverge at
infinity. Hence, they cannot be computed by simple basis state expansions.
(ii) In contrast to closed systems, a scattering system also contains a
continuous spectrum, associated with scattering solutions, which neither
diverge nor decrease in the asymptotic regions.
In numerical computations, the remnants of this continuous spectrum must be
separated from resonance eigenvalues.
(iii) The time-dependence of periodically-kicked scattering systems poses an
additional challenge.

\textit{Floquet theory} \cite{Flo1883, Sam1973} -- The time-periodic structure
of kicked-scattering systems results in resonance solutions which are
quasi-periodic in time.
The resonance eigenvalues, referred to as quasi-energies, may either be computed
from the so-called Floquet Hamiltonian or the stroboscopic time-evolution
operator.
Since the stroboscopic time-evolution operator is the quantum map, we here
choose the second approach.
Details are described in Sec.~\ref{Sec:Floquet}.

In order to face the divergence of resonance states we adapt two standard
methods to periodically-kicked scattering systems, namely, complex scaling
\cite{AguCom1971, BalCom1971, ReeSim1978, Rei1982b, Moi1998}, as previously
developed for time-periodic systems \cite{ChuRei1977, MaqChuRei1983, MoiKor1990,
BenMoiLefKos1991, BenMoiKosCer1993} and complex absorbing potentials, as
previously discussed for time-independent systems \cite{RisMey1993}:

\textit{Complex scaling} \cite{AguCom1971, BalCom1971, ReeSim1978} -- The main
idea of complex scaling is to consider the system along a complex contour
\begin{align}
  \label{eq:Contour}
   \mathcal{C}: s\in\mathbb{R} \mapsto q \in\mathbb{C} \quad \text{with:}
\quad q(s) := s\exp{\left(i \theta \right)},
\end{align}
determined by the scaling angle $\theta$.
While this turns the Schr{\"o}dinger equation into a non-Hermitian operator, it
also changes the asymptotic behavior of resonance states.
In particular, long-lived resonance states become square-integrable which makes
their eigenvalues amenable to numerical computations using basis state
expansions.
The method further rotates the continuous spectrum away from the real axis and
thus allows for its effective separation from resonance eigenvalues.
We emphasize that complex scaling is far more than just a numerical method to
compute resonances.
In particular, in the mathematical literature it is the method of choice for
dealing with resonances, see for example Ref.~\cite{ReeSim1978} for
time-independent systems, Refs.~\cite{Yaj1982, GraYaj1983, How1983} for
time-periodic systems and Ref.~\cite{NonSjoZwo2011} for open quantum maps.
An introductory description of complex scaling for time-independent systems is
given in Sec.~\ref{Sec:ComplexScalingTimeIndependent}.
The application of complex scaling to kicked scattering systems is discussed in
Sec.~\ref{Sec:ComplexScaling}.

\textit{Absorbing potentials} \cite{RisMey1993} -- Adding suitable absorbing
potentials to a scattering system represents an alternative method for computing
resonance spectra \cite{RisMey1993}.
The modified Hamiltonian generally neither supports resonance states nor a
continuous spectrum.
However, in the limit of weak absorption the point spectrum of the modified
Hamiltonian mimics the resonance spectrum and the continuous spectrum of the
original scattering system in a well understood manner \cite{RisMey1993}.
In that, resonance eigenvalues of a scattering system may be approximated by
numerically computed point spectra of absorption-augmented Hamiltonians.
The details of this method and its adaption to kicked-scattering systems are
discussed in Sec.~\ref{Sec:AbsorbingPotential}.

\subsection{Floquet theory and quantum maps}
\label{Sec:Floquet}

The Hamiltonian in Eq.~\eqref{eq:SchroedingerEquation} is time-periodic and we
have $\HOp(t)=\HOp(t+T)$ with $T=1$ and $\omega=2\pi/T$.
Its solutions admit a quasi-periodic representation \cite{Flo1883}
\begin{align}
  \label{eq:FloquetSolution}
  \!\!\!\ket{\psi_{\epsilon}(t)} = e^{-i\epsilon t / \hbar}\ket{u_{\epsilon}(t)}
\text{ with } \ket{u_{\epsilon}(t)} = \ket{u_{\epsilon}(t+1)},
\end{align}
with $\epsilon$ being a  quasi-energy.
The above decomposition is not unique.
In particular, replacing $\epsilon$ by $\epsilon+n\omega\hbar$ and
simultaneously multiplying $\ket{u_{\epsilon}(t)}$ by $\exp(in\omega t)$
yields identical solutions $\ket{\psi_{\epsilon}(t)}$ for any $n\in\mathbb{Z}$
\cite{Sam1973}.
In order to fix a unique representation of the solutions one can for example
shift the real parts of the quasi-energies to the first Floquet zone
$\myRe{(\epsilon)} \in [0,\omega\hbar)$.

In what follows we seek to determine the spectrum of quasi-energies via the
stroboscopic time-evolution operator $\qmap(t,t+T)$, henceforth referred to as
$\qmap$.
Applying this quantization of the classical map to a Floquet solution,
Eq.~\eqref{eq:FloquetSolution}, and suppressing the index $t$ gives
\begin{align}
  \label{eq:EigenvalueQmap}
  \qmap \ket{\psi_{\epsilon}} =
\exp{\left(-i\frac{\epsilon}{\hbar}\right)}\ket{\psi_{\epsilon}}.
\end{align}
Hence, the quasi-energies are related to the eigenvalues $\lambda_{\epsilon}$ of
the quantum map $\qmap$ as
\begin{align}
  \label{eq:Eigenvalue}
  \lambda_{\epsilon}:=
  \exp{\left(-i\frac{\epsilon}{\hbar}\right)}.
\end{align}

\subsection{Complex scaling for time-independent systems}
\label{Sec:ComplexScalingTimeIndependent}

In this section we briefly recall results on complex scaling of time-independent
systems as obtained in Refs.~\cite{AguCom1971, BalCom1971, BalCom1971,
ReeSim1978} and reviewed in Refs.~\cite{Rei1982b, Moi1998}.
This section is intended as an introduction for non-expert readers.

For the moment, consider a scattering system
\begin{align}
  \label{eq:TISE}
\left[E + \frac{\hbar^2}{2}\frac{\partial^2}{\partial q^{2}}
 - V(q) \right] \phi_{E}(q) = 0.
\end{align}
with a time-independent potential satisfying Eq.~\eqref{eq:DecayingPotential}.
Since the system is unbound, its spectrum contains resonance states, with
eigenvalues in the lower half of the complex plane $\myIm{(E)}<0$.

\begin{figure}[tb!]
  \begin{center}
    \includegraphics[]{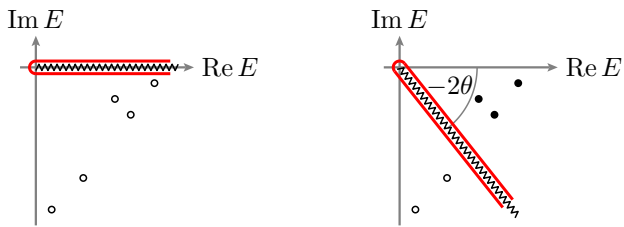}
     \caption{(color online) Schematic eigenvalue spectra (left) before and
(right) after complex scaling.
Resonance eigenvalues are shown as circles.
Open circles correspond to states, which are not square-integrable.
Closed circles correspond to states, which are square-integrable.
Continuous spectra are shown by lines.
Branch cuts of the complex energy root (zig-zag line) separate square-integrable
from non-square integrable solutions.
}
     \label{fig:spectra_TISE_schematic}
  \end{center}
\end{figure}

However, resonance states are not square-integrable.
To motivate this, consider the Schr{\"o}dinger equation in the region $q\gg1$
where $V(q)$ tends to zero.
Here, a solution takes the asymptotic form
\begin{align}
 \label{eq:AsymptoticSolutionTI}
 \phi_{E}^{\pm}(q) \simeq \exp{\left(\frac{i p^{\pm}(E)q}{\hbar}\right)},
\end{align}
with momenta
\begin{align}
 \label{eq:TIp}
 p^{\pm}(E) = \pm\sqrt{|2E|}\exp{\left(i\frac{\Arg{(E)}}{2}\right)}.
\end{align}
Here, the complex argument function $\Arg{(E)}$ maps to values on the interval
$(0,2\pi]$.
This fixes the branch cut of the energy root along the positive real axis, such
that:
(i) The first Riemann sheet $p^{+}$ gives square-integrable solutions.
On this sheet the Hamiltonian is hermitian and the corresponding spectrum must
be real.
(ii) The second Riemann sheet $p^{-}$ gives solutions which diverge as $q$
tends to infinity.
Here, one finds resonance solutions associated with eigenvalues in the lower
half of the complex plane.
(iii) The scattering solutions, which neither decay nor diverge as $q$ tends to
infinity, are along the branch cuts which separate both sheets.
They give the continuous spectrum along the positive real axis.
See Fig.~\ref{fig:spectra_TISE_schematic} for a sketch.

This motivates that a resonance wave function is not square integrable.
In the theory of complex scaling \cite{AguCom1971, BalCom1971, BalCom1971,
ReeSim1978} this problem is dealt with by considering wave functions along a
complex contour, Eq.~\eqref{eq:Contour}, resulting in the complex-scaled
wave-function
\begin{align}
  \label{eq:ScaledWaveFunctions}
  \phi_{E}^{\theta}(s) := \phi_{E}(q(s)).
\end{align}
and the complex-scaled Schr{\"o}dinger equation
\begin{align}
  \label{eq:TISECS}
\left[E
+\frac{e^{-i 2\theta }\hbar^2}{2}\frac{\partial^2}{\partial s^{2}}
 -V(se^{i\theta})  \right] \phi_{E}^{\theta}(s) = 0.
\end{align}
Here, $\phi_{E}^{\theta}(s)$ is from a suitable subspace of square-integrable
functions \cite{AguCom1971, BalCom1971, ReeSim1978}.
In contrast to common practice we do not multiply the wave-function by a
phase-factor $\exp{(i\theta/2)}$ as it has no effect on the resulting spectra.

The utility of this transformation is that a part of the resonance states
become square-integrable.
More precisely, if the potential $V(q)$ is dilation analytic \cite{ReeSim1978},
which implies that $V(q)$ has no singularity in the sector
$\Arg(q)\in[0,\theta]$ and further has the asymptotic behavior
\begin{align}
 \label{eq:DecayingPotentialCS}
\lim_{s\to\pm\infty}V(s\exp{i\alpha})=0 \quad  \forall\alpha\in[0,\theta],
\end{align}
we have:
(i) The point spectrum of the scaled Hamiltonian and the original Hamiltonian
are identical \cite{ReeSim1978}.
In particular, eigenvalues corresponding to resonance states are invariant under
variations of the scaling angle $\theta$ and those resonance states with
eigenvalues in the sector $\Arg(E) \in (-2\theta,0)$ become square integrable
along the complex contour, Eq.~\eqref{eq:Contour}.
(ii) The continuous spectrum of the scaled Hamiltonian is rotated into the
lower half of the complex plane along the ray $\mathbb{R}_{+}\exp{(-i2\theta)}$.

In order to motivate these properties, we consider the scaled equation in the
region $s\gg1$ where $V(s)$ tends to zero.
Here, the scaled form of the previously discussed asymptotic solutions reads
\begin{align}
 \label{eq:AsymptoticSolutionTICS}
 \phi_{E}^{\theta,\pm}(s)
 \simeq \exp{\left(\frac{i p^{\pm}(E)s \exp{(i\theta)}}{\hbar}\right)}.
\end{align}
Note that these are the same solutions as in Eq.~\eqref{eq:AsymptoticSolutionTI}
now considered along the contour, Eq.~\eqref{eq:Contour}.
This shows:
(i) Solutions in the sector $\Arg(E)\in(-2\theta,0)$ of the second Riemann
sheet $p^{-}(E)$, as previously defined in Eq.~\eqref{eq:TIp}, become
square-integrable when considered along the contour, Eq.~\eqref{eq:Contour}.
It is these resonances that complex scaling makes accessible to numerical
computations.
(ii) The branch cut which separates the sheet of square-integrable solutions
from the sheet of solutions which diverge as $s$ tends to infinity is now
located along the ray $\mathbb{R}_{+}\exp{(-i2\theta)}$.
This ray supports the scattering solutions of the scaled Hamiltonian which
neither decay nor diverge as $s$ tends to infinity.
It is along this ray that the scaled Hamiltonian has its continuous spectrum.
See Fig.~\ref{fig:spectra_TISE_schematic} for a sketch.

\subsection{Complex scaling for kicked scattering systems}
\label{Sec:ComplexScaling}

In this section we adapt the method of complex scaling, as previously developed
for periodically driven systems in Refs.~\cite{ChuRei1977, MaqChuRei1983,
Yaj1982, GraYaj1983, How1983, MoiKor1990, BenMoiLefKos1991, BenMoiKosCer1993},
to periodically-kicked scattering systems.
To this end, we first derive the complex-scaled Floquet Hamiltonian which is
convenient for discussing the structure of the resulting spectra.
We then derive an explicit expression for the complex-scaled time-evolution
operator, Eq.~\eqref{eq:PropagatorScaled}, i.e., the complex-scaled quantum map
and compute its spectrum, Fig.~\ref{fig:spectrum_CS}.

\subsubsection{Complex scaling for time-periodic Hamiltonians}

In order to apply complex scaling to time-periodic systems, we consider the
Schr{\"o}dinger equation~\eqref{eq:SchroedingerEquation} along the contour
Eq.~\eqref{eq:Contour}.
This gives wave-functions
\begin{align}
  \label{eq:ScaledWaveFunctions}
  \psi^{\theta}(s,t) := \psi(q(s),t),
\end{align}
and the complex-scaled Schr{\"o}dinger equation
\begin{align}
  \label{eq:SchroedingerEquationScaled}
\left[
i\hbar\frac{\partial}{\partial t}
+\frac{e^{-i 2\theta }\hbar^2}{2}\frac{\partial^2}{\partial
s^{2}}
- V(s e^{i \theta})\sum_{n\in\mathbb{Z}}\delta(t-n)
\right] \psi^{\theta}(s,t).
\end{align}
From Floquet theory, we infer the general form of a resonance state to admit
the Fourier representation
\begin{align}
 \label{eq:FloquetSolutionExpandedCS}
 \psi_{\epsilon}^{\theta}(s,t) = \sum_{\nu\in\mathbb{Z}}
\phi_{\epsilon,\nu}^{\theta}(s)e^{-i(\epsilon+\hbar\omega\nu)t/\hbar}.
\end{align}
Inserting this representation of eigenstates, into the scaled Schr{\"o}dinger
equation and integrating out the Fourier components results in a coupled
channel equation with channels $\mu\in\mathbb{Z}$.
For kicked scattering systems the equation of the $\mu$th channel reads
\begin{align}
 \label{eq:FloquetHamiltonianCS}
\left[
\epsilon+\mu\hbar\omega
+\frac{e^{-i 2\theta }\hbar^2}{2}\frac{\partial^2}{\partial
s^{2}} \right] \phi_{\epsilon,\mu}^{\theta}(s) =
V(s e^{i \theta})\sum_{\nu\in\mathbb{Z}}\phi_{\epsilon,\nu}^{\theta}(s).
\end{align}
Note that this equation is usually derived \cite{ChuRei1977, MaqChuRei1983,
MoiKor1990, Yaj1982, GraYaj1983, How1983} by first switching to the
time-independent Floquet Hamiltonian and then apply complex scaling
\cite{BalCom1971,ReeSim1978}.
While we will not use it for numerical computations, it is convenient for
discussing the structure of quasi-energy spectra in time-periodic systems as
done in the next section.
Once again we emphasize the crucial point of complex scaling:\
By considering the system along a complex contour we change the asymptotic
properties of resonance states.
In particular, long-lived resonance states $\psi_{\epsilon}^{\theta}(s,t)$ and
their Fourier components $\phi_{\epsilon,\nu}^{\theta}(s)$ become
square-integrable with respect to $s$.

\subsubsection{Structure of resonance spectra}

\begin{figure}[tb]
  \begin{center}
    \includegraphics[]{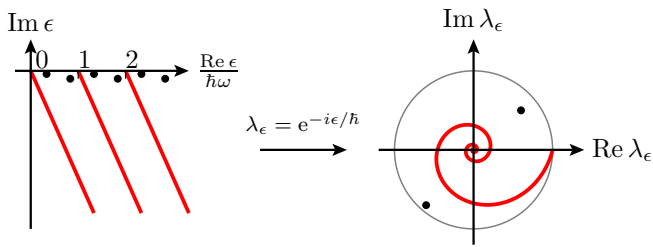}
     \caption{(color online) Schematic spectra of time-periodic scattering
systems, for the complex-scaled (left) Floquet Hamiltonian and (right)
time-evolution operator.
Resonance eigenvalues are shown as dots.
Continuous spectra are shown as (red) lines.
The gray line represents the unit circle.
}
     \label{fig:spectra_TPSE_schematic}
  \end{center}
\end{figure}

We now discuss the structure of the spectrum.
To this end we consider the Floquet Hamiltonian,
Eq.~\eqref{eq:FloquetHamiltonianCS}, in the asymptotic region
$s\gg1$ where the potential tends to zero.
This shows that the continuous spectrum of the $\mu$th channel is along the ray
\begin{align}
 \label{eq:ContinuousFloquet}
 \epsilon(b) = -\mu\hbar\omega + b\exp{(-i2\theta)}\quad\text{with: }
b\in\mathbb{R}_{+},
\end{align}
where, $-\mu\hbar\omega$ is the ionization threshold of the $\mu$th channel.
On the other hand, the theory of complex scaling ensures that resonance
eigenvalues in the lower half of the complex plane are invariant under
variations of the scaling angle \cite{AguCom1971, BalCom1971, BalCom1971,
ReeSim1978}.
See Fig.~\ref{fig:spectra_TPSE_schematic} for a sketch and
Ref.~\cite{MoiKor1990} for an example.

Note that the multi-channel structure of time-periodic system implies a new
feature.
Namely, increasing the scaling angle may not only uncovered new resonances.
In fact, a resonance eigenvalue may also be covered, if the continuous spectrum
of another channel sweeps over them \cite{Sim1973, Sim1978}.
This phenomenon and the corresponding structure of resonance states is
discussed in App.~\ref{App:StructureOfResonanceStates}.

The spectrum of the Floquet Hamiltonian, Eq.~\eqref{eq:FloquetHamiltonianCS},
is periodic under shifts by integer multiples of $\omega\hbar$.
This is because any solution $(\epsilon, \phi_{\epsilon,\mu}^{\theta}(s))$ gives
rise to a new solution $(\epsilon', \varphi_{\epsilon',\mu'}^{\theta}(s))=
(\epsilon+n\omega\hbar, \phi_{\epsilon,\mu'+n}^{\theta}(s))$ with shifted
eigenvalue.
Note that any of these shifted solutions gives rise to the identical Floquet
solution, Eq.~\eqref{eq:FloquetSolutionExpandedCS}, such that the full spectral
information is already stored in the first Floquet zone.
Here, we extend this zone into the complex domain along the rays of the
continuous spectrum
\begin{align}
 \label{eq:FloquetZone}
 \mathcal{F} :=
   \left\{\epsilon \in\mathbb{C} \,\,\,: \,\,\, \epsilon = a + b \,
e^{-i2\theta}, a \in [0,\hbar\omega), b \in \mathbb{R}_{+}
   \right\}.
\end{align}
As discussed in App.~\ref{App:StructureOfResonanceStates} a resonance state is
square-integrable and thus amenable to numerical computations, if the scaling
angle $\theta$ is chosen such that the quasi-energy $\epsilon$ associated with
the resonances lowest outgoing channel is located in the Floquet zone
$\epsilon\in\mathcal{F}$.

Rather than using the Floquet Hamiltonian, Eq.~\eqref{eq:FloquetHamiltonianCS},
we will follow the ideas of Refs.~\cite{BenMoiLefKos1991, BenMoiKosCer1993} and
compute the resonance spectrum via the eigenvalues of the stroboscopic
time-evolution operator.
As discussed in Sec.~\ref{Sec:Floquet}, both eigenvalues are related by
Eq.~\eqref{eq:Eigenvalue}.
This shows that both resonance eigenvalues and the continuous spectrum should
localize within the unit circle.
Furthermore, combining Eq.~\eqref{eq:Eigenvalue} and
Eq.~\eqref{eq:ContinuousFloquet} the continuous spectrum should localize along a
spiral, given by
\begin{align}
 \label{eq:ContinuumSpiral}
 \lambda(b) = \exp{\left(-i b \exp{\left(-i 2\theta \right)}\right)} \quad
\text{with: } b\in \mathbb{R}_{+}.
\end{align}
See Fig.~\ref{fig:spectra_TPSE_schematic} for a sketch.

\subsubsection{Complex-scaled time-evolution operator}

We now derive the complex-scaled time-evolution operator, i.e., the quantum map
from Eq.~\eqref{eq:SchroedingerEquationScaled}.
Clearly, in between two kicks the potential term is irrelevant and the particle
is subject to free motion in the complex-scaled Hamiltonian.
This gives the propagator of a free particle in its complex-scaled form
\begin{align}
  \label{eq:FreePropagatorScaled}
  \braOpket{s'}{\qmap_{VG}^{\theta}}{s} =
\frac{\exp{(i[\theta-\pi/4])}}{(2\pi\hbar)^{1/2}}
\exp{\left(\frac{i}{\heff}e^{i2\theta}\frac{(s-s')^2}{2}\right)}.
\end{align}
On the other hand, during the extremely short and intense pulse, the kinetic
energy term of Eq.~\eqref{eq:SchroedingerEquationScaled} is irrelevant and the
wave-function acquires a phase which is determined by the kicking potential
\begin{align}
  \label{eq:KickPropagatorScaled}
  \braOpket{s'}{\qmap_{HK}^{\theta}}{s} = \delta(s-s')
\exp{\left(-\frac{i}{\hbar} V\left(s e^{i\theta}\right)\right)}
\end{align}
Combining these operators, we obtain the complex-scaled version of the
time-evolution operator, Eq.~\eqref{eq:Qmap}, as
\begin{widetext}
\begin{align}
  \label{eq:PropagatorScaled}
  \braOpket{s'}{\qmap^{\theta}}{s} =
\frac{\exp{(i[\theta-\pi/4])}}{(2\pi\hbar)^{1/2}}
\exp{\left( -\frac{i}{2\hbar} V\left(s' e^{i\theta}\right)
            +\frac{ie^{i2\theta}}{\hbar} \frac{(s-s')^2}{2}
            -\frac{i}{2\hbar} V\left(s e^{i\theta}\right)
     \right)}.
\end{align}
\end{widetext}
This is the complex-scaled quantum map.
We emphasize that knowing the explicit form of the scaled time-evolution
operator is an advantage of kicked scattering systems over general
time-periodic systems
\cite{BenMoiLefKos1991, BenMoiKosCer1993}.

\subsubsection{Numerical spectra}
\label{Sec:ComplexScalingNumericalResults}

\begin{figure}[tb]
  \begin{center}
    \includegraphics[]{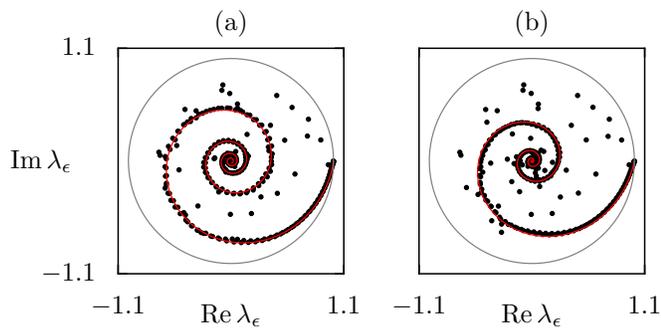}
     \caption{(color online) Numerically determined eigenvalues (black dots) of
the complex-scaled time-evolution operator, Eq.~\eqref{eq:PropagatorScaled},
for scaling angles (a) $\theta=0.05$ or (b) $\theta=0.08$.
The parameters of the model system are $\kappa=11$ and $\heff=1/50$.
The $\theta$-dependent continuous spectrum is along a spiral ([red] line).
The gray line represents the unit circle.}
     \label{fig:spectrum_CS}
  \end{center}
\end{figure}

We now discuss the computation of the resonance spectrum from the
complex-scaled time-evolution operator.
The numerical computations are carried out by expanding the operator in
Eq.~\eqref{eq:PropagatorScaled} on a basis set of square-integrable functions
and computing its eigenvalues.
In particular, our implementation follows Refs.~\cite{ScrEla1993, Scr2010}.
That is, we choose a finite element representation which allows for
representing square-integrable resonance wave functions with very high accuracy
and thus allows for computing highly accurate resonance eigenvalues.
Details are described in App.~\ref{App:NumericalImplementation}.
The results for the model system at kick-strength $\kappa=11$ and $\heff=1/50$
are illustrated in Fig.~\ref{fig:spectrum_CS}, for two values of the scaling
angle $\theta$.

As expected from Eq.~\eqref{eq:ContinuumSpiral}, the numerical remnants of the
continuous states localize along a spiral, marked by a red line in
Fig.~\ref{fig:spectrum_CS}.
Clearly, those states depend on the scaling angle $\theta$, giving two distinct
spiral configurations in Fig.~\ref{fig:spectrum_CS}(a) and (b).
On the other hand, we find stable $\theta$-independent eigenvalues away from
the spiral which correspond to resonance states.

We now make a couple of remarks:
(i) The resonance states, associated with eigenvalues which are sufficiently
away from the continuous spectrum, are exponentially localized along the
contour, Eq.~\eqref{eq:Contour}, (not shown).
For this reason, their eigenvalues are extremely stable not only under
variations of the scaling angle, but also under variations of other numerical
parameters, such as the length of the grid.
(ii) In contrast, the numerical eigenstates associated with the remnants of the
continuous spectrum do not decay towards the edges of the numerical grid.
For this reason their eigenvalues are affected by finite-size effects of the
numerical computation.
In particular, at large kick strength the numerical remnants of the continuous
spectrum do not localize exactly along the spiral predicted by
Eq.~\eqref{eq:ContinuumSpiral}.
For this reason the red lines in Fig.~\ref{fig:spectrum_CS} originate from
Eq.~\eqref{eq:ContinuumSpiral} where scaling angles have been enlarged over the
theoretically expected value by a factor of (a) $1.5$ and (b) $1.3$.
(iii) Note that, numerically computed resonance states are along the contour
$\mathcal{C}$.
Back-scaling such states to the real axis $q\in\mathbb{R}$ is numerically
unstable.
Therefore, the presented results cannot provide inside into the localization of
resonance states in real phase space.
This could be achieved using the method of exterior complex scaling
\cite{Moi1998}, and remains a future task.

\subsection{Resonance spectra from absorbing potentials}
\label{Sec:AbsorbingPotential}

The main purpose of this paper, namely the computation of resonance spectra for
periodically-kicked scattering systems has been achieved in the previous
section.
The main motivation for the rest of this paper is to make a connection to the
approach termed \textit{heuristic models} in the introduction.
Namely, we would like to discuss to which extend it is possible to obtain the
resonance spectra of periodically-kicked scattering systems, by adding
absorption to its time-evolution operator.

This section will start with a short review of Ref.~\cite{RisMey1993} in which
it was proven that adding a complex absorbing potential
\begin{align}
  \label{eq:AbsorbingPotential}
  V_{\eta}(q) &= -i \eta v(q),
\end{align}
to a time-independent Hamiltonian and letting the absorption strength
$\eta\in\mathbb{R}_{+}$ tend to zero ($\eta\to0$), allows for computing the
resonance spectrum.
We will then adapt the method of absorbing potentials to periodically-kicked
scattering systems in a rather heuristic manner, namely by adding an absorbing
potential to the kick.
This will result in an absorption-augmented time-evolution operator,
Eq.~\eqref{eq:PropagatorAbsorption}, which takes exactly the form of heuristic
model systems.
We then demonstrate that the spectra of the absorption-augmented time-evolution
operator indeed recover the resonance spectra of the model system in the limit
of weak absorption ($\eta\ll1$), see Fig.~\ref{fig:spectrum_VABS}.
In contrast, we also show that the resonance spectra of the model system cannot
be recovered, if applying projective openings in the outgoing regions.
Finally, we remark that the notion of absorbing potential, as used in this
paper, is distinct from any sort of exterior complex scaling, which from time
to time is also referred to as an absorbing potential in the literature.

\subsubsection{Time-independent systems}
\label{Sec:MainIdeaOfAbsorbingPotentials}

Computing resonance spectra of time-independent Hamiltonian systems based on
complex absorbing potentials, Eq.~\eqref{eq:AbsorbingPotential}, has been
established in Ref.~\cite{RisMey1993}.
The main idea of the method is quite different from complex scaling in the
following sense:\
While complex scaling considers the same system along a different contour in
coordinate space, the method of absorbing potentials adds a complex absorbing
potential to the scattering system.
As previously emphasized in Sec.~\ref{Sec:Overview} this changes the properties
of the system entirely.
In particular, the absorption-augmented scattering system is usually closed and
exhibits neither a continuous spectrum nor resonance states.
However, in the limit of weak absorption $(\eta\to0)$ the point spectrum of the
modified system approximates the resonance eigenvalues of the original system
in a well controlled manner.

More specifically, the method of absorbing potentials \cite{RisMey1993}, adds
an absorbing potential, Eq.~\eqref{eq:AbsorbingPotential}, to the
Schr{\"o}dinger equation.
As specified in Ref.~\cite{RisMey1993} one may use a wide class of potentials
$v(q)$ which fulfill certain properties.
In particular, in the asymptotic regions we require the real part of $v(q)$ to
tend to infinity.
Here, we assume $v(q)$ which fulfill the conditions spelled out in
Ref.~\cite{RisMey1993} with asymptotic forms of a monomial
\begin{align}
 v(q)\simeq|q|^n \quad\text{for: } |q|\gg1, n>0.
\end{align}

In that case, the results of Ref.~\cite{RisMey1993} state that, adding an
absorbing potential, Eq.~\eqref{eq:AbsorbingPotential}, to a time-independent
Hamiltonian, Eq.~\eqref{eq:TISE}, and considering the limit of weak absorption
($\eta\to0$), the spectra of the absorption-augmented Hamiltonians will converge
onto the following shape:\
(i) The remnants of the continuous spectrum (referred to as spectral string in
Ref.~\cite{RisMey1993}) will arrange along the ray
$\mathbb{R}_{+}\exp{(-i2\bar{\theta})}$, with
\begin{align}
 \bar{\theta}=\frac{\pi}{2n+4}.
\end{align}
(ii) For each resonance solution of Eq.~\eqref{eq:TISE} with energy $E$ in the
sector $\Arg(E)\in(-2\bar{\theta},0)$, the family of absorption-augmented
Hamiltonians gives a family of eigenvalues $E(\eta)$ which tends to the
resonance eigenvalue $E$ of the scattering system as $\eta$ tends to zero,
according to
\begin{align}
 \lim_{\eta\to0}E(\eta)=E.
\end{align}

We now make a couple of remarks:\
(i) The structure of the spectrum and in particular the role of the angle
$\bar{\theta}$ is reminiscent of complex scaling.
(ii) In numerical computations the strength of the absorbing potential must be
reduced successively, while simultaneously monitoring the convergence of the
corresponding spectra.
(iii) The intuitive interpretation of the complex absorbing potential method is
as follows:\
In order to compute the resonance spectrum, we add an absorbing potential which
turns the asymptotic, exponentially diverging and outgoing parts of a resonance
state into a square-integrable solution.
However, by adding an absorbing potential, we introduce unphysical
back-reflections from the absorbing potential itself, which deteriorate the
quality of the solution.
In order to minimize this unphysical back-reflection, we have to consider the
limit of weak absorption ($\eta\ll1$) in which the quality of the solution
improves.

\subsubsection{Absorption-augmented time-evolution operator}

We now adapt the results of Ref.~\cite{RisMey1993} to periodically-kicked
scattering systems in an add-hoc manner.
That is, rather than adding the complex absorbing potential to the
Schr{\"o}dinger equation, we add it directly to the kicking potential
\begin{align}
  \label{eq:SchroedingerEquationAbsorption}
\left[
 i\hbar\frac{\partial}{\partial t}
 \!+\!\frac{\hbar^2}{2}\frac{\partial^2}{\partial q^{2}}
 \!-\![V(q)\!+\!V_{\eta}(q)]\sum_{n\in\mathbb{Z}}\!\delta(t\!-\!n)
\right] \!\psi(q,t) =0.
\end{align}
The motivation for doing so is twofold:\
(i) Grouping the complex absorbing potential with the kick potential allows for
a straight-forward computation of the corresponding time-evolution operator.
(ii) Our physical intuition is that grouping the complex absorbing potential
with the kick, will do an equally good job in transforming the asymptotic parts
of a resonance solution into square-integrable functions and thus allows for
computing the resonance spectrum.

As in the previous section, we will access the quasi-energy spectrum of
Eq.~\eqref{eq:SchroedingerEquationAbsorption} via the corresponding
time-evolution operator.
This time-evolution operator is easily obtained by taking the time-evolution
operator of the previous section, Eq.~\eqref{eq:PropagatorScaled}, considering
it for scaling angle $\theta=0$, relabeling $s$ by the standard variable $q$,
and replacing $V(q)$ by $V(q)-i \eta v(q)$.
This gives the absorption-augmented time-evolution operator according to
\begin{widetext}
\begin{align}
  \label{eq:PropagatorAbsorption}
  \braOpket{q'}{\qmap^{\eta}}{q} = P^{\eta}(q) \times
\frac{\exp{(-i\pi/4)}}{(2\pi\hbar)^{1/2}}
\exp{\left( -\frac{i}{2\hbar} V\left(q'\right)
            +\frac{i}{\hbar} \frac{(q-q')^2}{2}
            -\frac{i}{2\hbar} V\left(q\right)
     \right)} \times P^{\eta}(q).
\end{align}
\end{widetext}
Here, the complex absorbing potential has been split from the standard kicking
potential.
This gives the absorption-augmented time-evolution operator as a composition of
the standard unitary time-evolution operator of the kicked scattering system,
multiplied with an absorption operator
\begin{align}
  \label{eq:AbsorptionOperator}
 P^{\eta}(q) = \exp{ \left(-\frac{\eta}{2\hbar}v(q)\right)}
\end{align}
from its left and its right.
Note that Eq.~\eqref{eq:PropagatorAbsorption} takes the typical form of a
system, termed \textit{heuristic model} in the introduction.

We will now explore under which condition absorbing potentials allow for
computing the resonance spectrum of a kicked scattering system.
To this end, we consider the model system of this paper and compute the
corresponding spectrum of Eq.~\eqref{eq:PropagatorAbsorption} for weak and
strong absorbing potentials as well as projective openings.
This will show that only for weak absorbing potentials the resonance spectrum
of Eq.~\eqref{eq:PropagatorAbsorption} coincides with the result obtained from
complex scaling.

\subsubsection{Weak absorbing potentials}

In order to adapt the method of Ref.~\cite{RisMey1993} and check it for the
model system of this paper, we fix $v(q)$ such that
\begin{align}
  v(q) &= \left(M_{n}^{\width}(q-\qabs) + M_{n}^{\width}(-q-\qabs)\right).
\end{align}
Here, $M_{n}^{\width}(x)$ is a smooth function which approaches a monomial,
when taking its smoothness parameter $\width$ to zero as
\begin{align}
 \lim_{\width\to0} M_{n}^{\width}(x) = \begin{cases}
                                       0       & \quad \text{if } x < 0\\
                                       x^{n}   & \quad \text{if } x \ge 0.\\
                                     \end{cases}
\end{align}
For details see App.~\ref{App:SmoothMonomials}.
This gives an absorbing potential which is close to zero for
$q\in(-\qabs,\qabs)$ such that the scattering region is unaffected by
absorption.

In order to numerically determine the spectrum of $\qmap^{\eta}$,
Eq.~\eqref{eq:PropagatorAbsorption}, we expand the corresponding modes.
To this end we choose a finite element representation on a grid with
$q\in[-\qmax,\qmax]$, see App.~\ref{App:NumericalImplementation} for details.

In numerical computations we implement the limit of weak absorption
($\eta\ll1$) by choosing the absorption strength as a function of the grid
parameter $\qmax$
\begin{align}
  \label{eq:AbsorptionStrength}
 \eta(\qmax) = -\frac{2 \hbar}{v(\qmax)} \log(10^{-15}).
\end{align}
This choice ensures that the absorption operator, defined in
Eq.~\eqref{eq:AbsorptionOperator}, fulfills
\begin{align}
  P^{\eta}(q) < 10^{-15} \quad \forall |q|>\qmax,
\end{align}
such that, any probability which is propagated beyond the grid edges by the
unitary part in Eq.~\eqref{eq:PropagatorAbsorption}, will be reduced below
machine precision.
The expert reader might notice that this eliminates the error of
back-reflections from the grid edges within the numerical desired accuracy.
We then minimize the back-reflections from the absorbing potential itself.
To this end we reduce the absorption strength $\eta$ successively by increasing
the grid size $\qmax$ via Eq.~\eqref{eq:AbsorptionStrength}.
Simultaneously we monitor the convergence of the spectra.

\begin{figure}[tb]
  \begin{center}
    \includegraphics[]{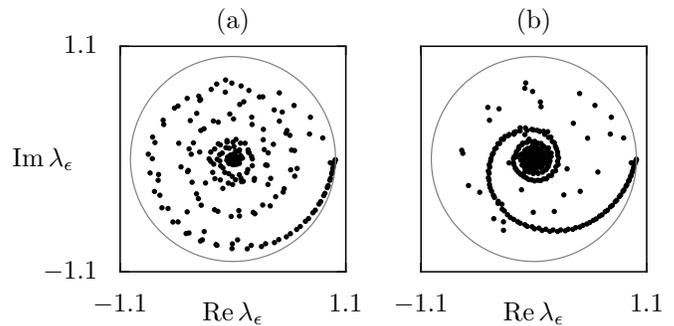}
     \caption{(color online) Numerically determined eigenvalues (black dots) of
the time-evolution operator, Eq.~\eqref{eq:PropagatorAbsorption} with (a)
strong and (b) weak absorption, for $\kappa=11$ and $\heff=1/50$.
The parameters of the absorbing potential are $\qabs=2.5$, monomial order
$n=3$, and width parameter $\width=0.1$.
The absorption strength $\eta$ is determined by the length of the numerical
grid $\qmax$ according to Eq.~\eqref{eq:AbsorptionStrength} as (a) $\qmax=2.51$
(strong absorption) and (b) $\qmax=5.0$ (weak absorption).
The unit circle is marked by a gray line.}
     \label{fig:spectrum_VABS}
  \end{center}
\end{figure}

The results are depicted in Fig.~\ref{fig:spectrum_VABS} where (a) shows a
spectrum for which the absorption strength was large while (b) shows a spectrum
for which the absorption strength was small.
These results show:\
(i) The agreement between the spectrum in Fig.~\ref{fig:spectrum_VABS}(b) with
the spectra obtained from complex scaling, Fig.~\ref{fig:spectrum_CS} is
striking.
(ii) We expect that the spiral structure in Fig.~\ref{fig:spectrum_VABS}(b)
corresponds to the remnants of the continuous spectrum.
This expectation is supported by the structural analogy between spectra of
time-independent systems, obtained from complex scaling on the one hand and
absorbing potentials on the other hand.
(iii) We further observe that the eigenvalues away from the spiral in
Fig.~\ref{fig:spectrum_VABS}(b), localize in exactly the same positions as the
resonance spectrum obtained from complex scaling in Fig.~\ref{fig:spectrum_CS}.
These eigenvalues should represent the resonance spectrum.
(iv) On the other hand, the spectrum in Fig.~\ref{fig:spectrum_VABS}(a), as
obtained for strong absorption, has little resemblance with the results obtained
from complex scaling, Fig.~\ref{fig:spectrum_CS}.
In particular, the continuous spectrum blows up and the separation between
resonance spectra and continuous spectra is lost.
We expect that this is due to unphysical reflections from the absorption
operator which deteriorate the quality of the solution.
In conclusion, we believe that the resonance spectrum of periodically-kicked
scattering systems can only be obtained from the absorption-augmented
time-evolution operators in the limit of weak absorption.

Finally, we speculate on the generality of the method.
While the results of this section have been obtained for one specific model
system, we are positive that the method of weak absorbing potentials should
allow for computing resonance spectra of any periodically-kicked scattering
system.
We believe so, since the absorbing potential chosen here is non-zero only in the
outgoing regions.
Here, its only task is to tweak the exponentially diverging tales of the
resonance wave-functions into square-integrable functions.
However, this structure is universally the same for any kicked scattering
system.
Nonetheless, putting the method on an equally solid mathematical footing as in
the case of time-independent systems \cite{RisMey1993} or in the case of
complex scaling, as discussed in the previous section, remains a future task.

\subsubsection{Projective opening}
\label{Sec:ProjectiveOpening}

Finally, we consider the case of projective openings, as commonly used for
heuristic model systems and show that they do not reproduce the resonance
spectrum of a kicked scattering system.

\begin{figure}[tb]
  \begin{center}
    \includegraphics[]{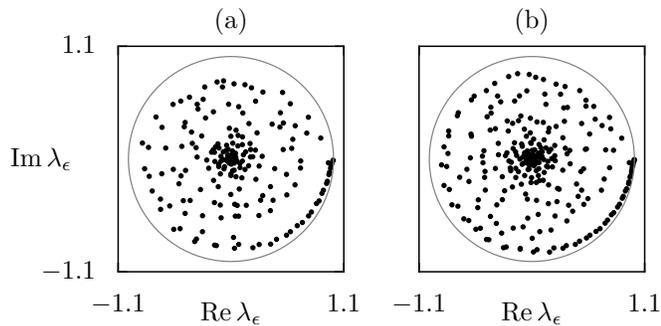}
     \caption{(color online)  Numerically determined eigenvalues (black dots) of
the time-evolution operator with projective openings for $\kappa=11$ and
$\heff=1/50$.
The projector removes probability for positions $|q|>\qabs$, where (a)
$\qabs=2.0$ and (b) $\qabs=2.5$.
The unit circle is marked by a gray line.}
     \label{fig:SpectrumBox}
  \end{center}
\end{figure}

To this end we consider an absorbing potential, which is zero for
$q\in(-\qabs,\qabs)$ and infinitely large outside of this interval.
This turns the absorption operator, Eq.~\eqref{eq:AbsorptionOperator}, into a
projector
\begin{align}
 \label{eq:Projector}
 P^{\eta}(q) := \begin{cases}
            1       & \quad \text{if } q \in  (-\qabs,\qabs)\\
            0       & \quad \text{if } q \notin (-\qabs,\qabs).\\
          \end{cases}
\end{align}
Here, we choose sufficiently large $\qabs$, such that the scattering region and
the trapped set in particular, are unaffected.
Based on this setting we consider the absorption-augmented time-evolution
operator, Eq.~\eqref{eq:PropagatorAbsorption}, and compute its spectrum
numerically.
See App.~\ref{App:NumericalImplementation} for details of the numerical
computation.
The result is depicted in Fig.~\ref{fig:SpectrumBox} for $\qabs=2.0, 2.5$.

These results show:
(i) Spectra obtained with projective openings have little resemblance with the
resonance spectrum obtained from complex scaling, see
Fig.~\ref{fig:spectrum_CS}, or weak absorbing potentials, see
Fig.~\ref{fig:spectrum_VABS}(b).
(ii) Similar to the case of strong absorption, Fig.~\ref{fig:spectrum_VABS}(a),
we observe that the continuous spectrum blows up.
In particular, the clear separation of the continuous spectrum and the
resonance spectrum is lost.
We speculate that the origin of this problem is again due to unphysical
back-reflections into the scattering region which originate from the
non-analytic corners of the absorbing potential.
We conclude that resonance spectra of kicked scattering systems are not
computable by virtue of projective openings.
(iii) Our results also have important implications for numerical tests of the
fractal Weyl law, based on heuristic model systems.
Namely, if we were to follow the standard recipe of heuristic models and
construct an open quantum map for the model system of this paper, we would apply
projective openings in the outgoing region far away from the trapped set.
However, our results show that the eigenvalues of this system have no relation
to the true resonance eigenvalues of the kicked scattering system, as obtained
from complex scaling or sufficiently weak absorbing potentials.
Hence, counting long-lived eigenvalues of heuristic model systems may not be a
reasonable test of fractal Weyl conjectures.

\section{Summary and discussion}
\label{Sec:SummaryAndDiscussion}

\subsection{Summary of the objective}

Ever since their invention \cite{BerBalTabVor1979, CasChiIzrFor1979,
HanBer1980}, quantum maps have been popular toy models for investigations of
quantum-to-classical correspondence.
So far, they have often been considered on compact phase spaces
\cite{HanBer1980, DegGra2003}, rendering their time-evolution operators as
finite-dimensional unitary matrices.
For this reason quantum maps are typically associated with closed systems.
However, studies of chaotic scattering \cite{Eck1987, GasRic1989a, GasRic1989b,
GasRic1989c, BluSmi1989, BluSmi1990, Jun1990, JunPot1990, JunTel1991,
CviEck1989, TanRicRos2000} and investigations on fractal Weyl laws
\cite{Sjo1990, Lin2002, LuSriZwo2003, SchTwo2004, Non2011, Nov2013,
KoeMicBaeKet2013}  and localization of resonance states \cite{KeaNovPraSie2006,
KoeBaeKet2015}, in particular, made an extension of these model systems to
quantizations of an open flow desirable \cite{NonSjoZwo2011}.
These model systems are known in the literature as open quantum maps, see
Ref.~\cite{NonSjoZwo2011} for their definition, Refs.~\cite{Non2011, Nov2013}
for a review, and Refs.~\cite{CasMasShe1997, CasGuaMas2000, SchTwo2004,
NonZwo2005, KeaNovPraSie2006, NonZwo2007, NonRub2007, KeaNovSch2008,
KeaNonNovSie2008, NovPedWisCarKea2009, KopSch2010, Non2011, IshAkaShuSch2012,
LipRyuLeeKim2012, Nov2013, KoeMicBaeKet2013, SchAlt2015, KoeBaeKet2015,
MerKulLoeBaeKet2016, FriBaeKetMer2017} for examples.

As discussed in the introduction, currently known examples of open quantum maps
- termed \textit{heuristic models} throughout this paper - exhibit two aspects,
which are disadvantageous from our point of view:\
(i) The first aspect is the add-hoc creation of concrete example systems by
simply combining a unitary matrix with an absorption operator.
This is confusing because it disguises the connection of a quantum map with a
concrete scattering system and further gives the impression that absorption
must be added to the system in order to introduce resonances.
(ii) The second disadvantageous aspect of current model systems is the use of
absorption operators with non-analytic parts, mainly projectors.
This introduces diffraction to the system and thus corrupts the utility of
quantum maps for pure investigations of quantum-to-classical correspondence.
It is for these two reasons that we propose a new approach to quantum maps with
resonances in this paper.

\subsection{Summary of the results}

In this paper, we consider quantum maps, induced by periodically-kicked
scattering systems.
In order to compute their resonance spectra, we apply the method of complex
scaling \cite{AguCom1971, BalCom1971, ReeSim1978, Rei1982b, Moi1998} as
previously developed for time-periodic systems \cite{ChuRei1977, MaqChuRei1983,
MoiKor1990, BenMoiLefKos1991, BenMoiKosCer1993}.
This results in a complex-scaled time-evolution operator,
Eq.~\eqref{eq:PropagatorScaled}, i.e., the \textit{complex-scaled quantum map}
from which the resonance spectra, Fig.~\ref{fig:spectrum_CS} or
Fig.~\ref{fig:spectrum_box_VABS_CS}(a), can readily be computed.

In order to examine the relation between the complex-scaled quantum map and
heuristic models, we further adapt the method of absorbing potentials
\cite{RisMey1993} to periodically-kicked scattering systems.
This results in an absorption-augmented time-evolution operator,
Eq.~\eqref{eq:PropagatorAbsorption}, which takes the form of heuristic models.
We show that this operator also allows for computing resonance spectra, in the
limit of sufficiently weak absorption, Fig.~\ref{fig:spectrum_VABS}(b) or
Fig.~\ref{fig:spectrum_box_VABS_CS}(b).
On the other hand, neither strong absorption, Fig.~\ref{fig:spectrum_VABS}(a) or
Fig.~\ref{fig:spectrum_box_VABS_CS}(c), nor projective openings,
Fig.~\ref{fig:SpectrumBox} or Fig.~\ref{fig:spectrum_box_VABS_CS}(d), as
commonly used for heuristic models, allow for computing resonance spectra of
kicked scattering systems.
Note that this is even the case, if the projective opening or the absorbing
potential do not affect the trapped set, which carries the semiclassical
information on the resonance spectrum.

Both for complex scaling and weak absorbing potentials a novel point, as
compared to heuristic models, is the appearance of a continuous spectrum, which
must be carefully separated from the resonance spectrum.
We show that this separation cannot be accomplished when using the
absorption-augmented time-evolution operator in the case of strong absorption,
Fig.~\ref{fig:spectrum_VABS}(a) or Fig.~\ref{fig:spectrum_box_VABS_CS}(c), and
projective openings, Fig.~\ref{fig:SpectrumBox} or
Fig.~\ref{fig:spectrum_box_VABS_CS}(d).
In both cases the numerically determined spectra exhibit a blow up of the
continuous spectrum which obscures its effective separation from the resonance
spectrum.

In order to support our claim with numerically computed spectra, we introduce a
model system, based on a family of analytic kicking potentials and discuss the
organization of its classical flow in terms of few stable and unstable
manifolds \cite{MacMeiPer1984, MacMeiPer1984b, Rom1990, Wig1992, Rom1994}.
In that, we emphasize classical structures of the model system which commonly
appear in classical ionization \cite{MitHanTigFloDel2004, MitHanTigFloDel2004b,
BurMitWykYeDun2011} and dissociation \cite{DavGra1986}.
Moreover, we demonstrate that the trapped set of the model system is organized
by a topological horseshoe at large kick strength.

\subsection{Discussion}

In this paper we exploit the stroboscopic time-evolution operators of
periodically-kicked scattering system in order to construct quantum maps which
posses resonances and have a clear origin in a scattering system.
Surprisingly, the resulting time-evolution operator is both a \textit{quantum
map} in the original sense of Ref.~\cite{BerBalTabVor1979, CasChiIzrFor1979} as
well as an \textit{open quantum map} in the sense of Ref.~\cite{NonSjoZwo2011},
albeit not yet reduced to a finite-dimensional matrix.

The main result and purpose of this paper was to show that the strategy of
heuristic model systems, i.e., combining an arbitrary unitary time-evolution
operator with an arbitrary absorption operator, such as a projector, cannot
in general be expected to produce the correct resonance spectrum of a scattering
system.
This fact is well established in the quantum chemistry community, where either
complex scaling \cite{Moi1998} or weak absorbing potentials \cite{RisMey1993}
are the established ways to compute resonance spectra.
In this paper we merely repeat this result by adapting it to quantum maps
originating from kicked scattering systems.
In that, our results provide a clear alternative to heuristic models and
represent a mathematically safe way to discuss concrete examples of quantum
maps with resonance spectra.

Further than that, we propose a kicked scattering system based on kick
potentials which are entire functions.
Hence, in contrast to many heuristic models which make use of absorption
operators with non-analytic part, such as projectors, the model system
discussed in this paper should be free from diffraction.
Hence, using them should be advantageous when studying matters of
quantum-to-classical correspondence.

As compared to heuristic models, the benefits of a kicked scattering system,
come at the price of a slightly increased effort in their numerical treatment.
Nonetheless, the numerical effort should still be lower than treating more
advanced systems, such as two-dimensional potential systems
\cite{RamPraBorFar2009}, three-disc scatterers \cite{GasRic1989c}, or atomic
systems \cite{TanRicRos2000}.
In particular, discussing small wave-length limits with effective Planck
constants as small as $\heff=1/100$ is still possible on a desktop PC.

Nonetheless, while we do believe that quantum maps as discussed in this paper
have several clear advantages over heuristic models, we do not think that
results obtained from heuristic models are generally wrong.
On the contrary!
For example, combining a cavities transfer operator \cite{Bog1992} with
Fresnel-type reflection, as described in Ref.~\cite{KeaNovSch2008} should give
an excellent open quantum map model of a dielectric cavity.
In our opinion, the results of this paper indicate that the key question in
connecting scattering systems on the one hand and their reduction to a
finite-dimensional open quantum map on the other hand, lies in effectively and
correctly removing the continuous spectrum.
With respect to this question, it seems reasonable to distinguish two classes
of scattering systems, as previously proposed in Ref.~\cite{AltPorTel2013}:\
(i) On the one hand, there are \textit{leaky scattering systems}
\cite{AltPorTel2013} which are obtained by opening a well-defined closed system.
Examples of such systems are quantum dots with leads or dielectric cavities.
For such systems it is straight forward to specify a subspace associated with
resonance states and Feshbach-type arguments \cite{Fes1958,Fes1962,Rot2009}
should be efficient in removing the continuous spectrum and deriving an
effective Hamiltonian.
Such systems should be well described by heuristic models.
(ii) On the other hand, there are \textit{ionizing scattering systems} which
are not related to any kind of closed system.
An examples of such a system is Helium.
In that case, it may not be obvious how to decompose the system by Feshbach-type
arguments and a safe and established way to access its resonances is provided
by means of complex scaling or absorbing potentials.
The kicked scattering systems of this paper and their quantum maps should be
suitable models for this class of systems.

Ultimately, it should also be possible to reduce such ionizing systems to
finite-dimensional sub-unitary matrices.
This could be achieved by implementing the procedure of
Ref.~\cite{NonSjoZwo2011} for a concrete system or by putting the idea of
Ref.~\cite{NovPedWisCarKea2009} to a test in a kicked scattering system.

\subsection{Future directions}

The model system proposed in this paper is an excellent model of ionization and
should be immediately useful for future investigations of dynamical tunneling in
systems with a mixed phase space as well as further investigations of fractal
Weyl laws.
For the latter, it would be useful, if the uniform hyperbolicity of the
proposed model system could be proven.
In contrast to the three-disc scatterer and the open backer map, this would
result in a uniformly hyperbolic model system which is not affected by
diffraction.
Furthermore, a proof of uniform hyperbolicity would also be a firm basis for
more detailed investigations of the model systems classical dynamics, e.g., in
terms of Lyapunov exponents, escape rates, entropies, and fractal dimensions
\cite{GasRic1989a, EckRue1985}.

We further believe that the model system is a suitable candidate for
investigating the localization of resonance states on classical invariant sets
and their implications for fractal Weyl laws in more detail.
In particular, the results of Ref.~\cite{SchTwo2004, KeaNonNovSie2008} are
entirely based on the existence of a leak.
The complement of its preimages down to the Ehrenfest time is used to specify
the support of long-lived resonances.
However, such a leak does not exist in ionizing scattering systems, such that
understanding the localization of their resonance states remains an open
problem.
Our model system should be ideal for addressing this question.
To this end, it is crucial to make exterior complex scaling available for
kicked scattering systems, which remains a future task.

Finally, an interesting direction of future research concerns the scattering
matrix of kicked scattering systems.
In particular, we wonder to which degree the structure of a periodically-kicked
scattering system may allow for explicitly evaluating several formulas,
described in Refs.~\cite{PesMoi1994, Moi1998}.
This would be useful for providing an outside view on open quantum maps and
could lead to further interesting results.

\begin{acknowledgments}

It is our great pleasure to gratefully acknowledge fruitful discussions with
Eduardo G. Altmann,
Arnd B{\"a}cker,
Stephen Creagh,
Carl Dettmann,
Italo Guarneri,
Yasutaka Hanada,
Takahisa Harayama,
Jon Keating,
Roland Ketzmerick,
Martin K\"orber,
Tamiki Komatsuzaki,
Jizhou Li,
Domenico Lippolis,
St{\'e}phane Nonnenmacher,
Stefan Rotter,
Yuzuru Sato,
Sulimon Sattari,
Susumu Shinohara,
Henning Schomerus,
Martin Sieber,
Atsushi Tanaka,
Gregor Tannor,
Hiroshi Teramoto,
Mikito Toda,
and
Stephen Wiggins.
N.M.\ acknowledges financial support by Deutsche Forschungsgemeinschaft (DFG)
via Grant No.\ ME 4587/1-1.
\end{acknowledgments}

\begin{appendix}

\section{Outgoing and incoming region}
\label{App:EscapeRegion}

In this section we show that a system with kicking potential as defined in
Eqs.~\eqref{eq:KickingPotential} -- \eqref{eq:ParameterConditions}, gives rise
to incoming and outgoing regions, as defined in
Eqs.~\eqref{eq:OutgoingRegionPlus} -- \eqref{eq:IncomingRegionMinus}.

We start by showing that an outgoing region $\mathcal{O}^{+}$ is ensured in
general, if
\begin{align}
  \label{eq:NegativityOfKick}
  V'(q)<0 \quad \forall q>\xf.
\end{align}
To demonstrate this claim we exploit condition \eqref{eq:NegativityOfKick} in
\eqref{eq:HalfKickPos} to show that for all $(q_{n}, p_{n})\in\mathcal{O}^{+}$,
i.e., $\forall (q_n, p_n)$ such that $q_n>\xf$ and $p_n>0$, we have
\begin{align}
  \label{eq:GrowthOfPosition}
  q_{n+1}&>q_{n} + p_{n}>q_{n}>\xf,
\end{align}
i.e., for points in $\mathcal{O}^{+}$ the position grows in one step of the
iteration.
Further, exploiting Eqs.~\eqref{eq:NegativityOfKick} and
\eqref{eq:GrowthOfPosition} in \eqref{eq:HalfKickMom} shows that for all
$(q_{n}, p_{n})\in\mathcal{O}^{+}$ we have
\begin{align}
  p_{n+1}&>p_{n}>0,
\end{align}
i.e., for points in $\mathcal{O}^{+}$ the momentum also grows in one step of the
iteration.
This means that for trajectories which enter $\mathcal{O}^{+}$ the position and
momentum is monotonically increasing such that trajectories which enter
$\mathcal{O}^{+}$ remain in $\mathcal{O}^{+}$.
Even more, the above equations imply that, if $(q_{n},
p_{n})\in\mathcal{O}^{+}$, we have that future positions $q_{m}$ with $m>n$
grow at least as
\begin{align}
  \label{eq:GrowthOfPositionII}
  q_{m+n}&>q_{n} + (m-n) p_{n},
\end{align}
which shows that any trajectory entering the region $\mathcal{O}^{+}$
will escape to infinity as
\begin{align}
 \lim_{m\to\infty}q_{m}\to\infty.
\end{align}
Note that for the model system in this paper the related properties for the
regions $\mathcal{O}^{-}$, $\mathcal{I}^{+}$, and $\mathcal{I}^{-}$ follow due
to parity and time-reversal symmetry.

We now show that condition~\eqref{eq:NegativityOfKick} holds for the model
system of this paper.
To this end we compute the derivate of the kicking potential,
Eq.~\eqref{eq:KickingPotential}, whose parts are defined in
Eqs.~\eqref{eq:GaussianKick} and \eqref{eq:PerturbationKick} using the
perturbation strength as defined in Eq.~\eqref{eq:PerturbationStrength}.
The resulting derivative is
\begin{align}
  &V'(q) = \kappa \, \exp{\left(-8q^{2}\right)} \times f(q),
\end{align}
with
\begin{align}
  &f(q) = q - \xf\frac{\exp(16\xb[q-\xf]) - \exp(-16\xb[q+\xf])}{1
- \exp\left(-32\xf\xb\right)}.
\end{align}
Throughout this paper we choose $\kappa>0$ which implies that $\kappa \,
\exp{\left(-8q^{2}\right)}>0$.
Under this condition showing that condition~\eqref{eq:NegativityOfKick} holds is
equivalent to showing that
\begin{align}
  \label{eq:NegativityOfKickf}
  f(q)<0 \quad \forall q>\xf.
\end{align}
We show that condition~\eqref{eq:NegativityOfKickf} holds from the fact that
$f(\xf)=0$ while $f(x)$ is strictly monotonically decreasing for $q\geq\xf$.
The latter can be shown from the derivative
\begin{align}
  f'(q) = 1 - 16\xb\xf\frac{\exp(16\xb[q-\xf]) + \exp(-16\xb[q+\xf])}{1
- \exp\left(-32\xf\xb\right)},
\end{align}
being strictly negative
\begin{align}
  f'(q)<0 \quad \forall q\geq\xf.
\end{align}
The last condition is ensured by Eq.~\eqref{eq:ParameterConditions}, and the
fact that $q\ge\xf$ implies that $\exp(16\xb[q-\xf])\ge1$,
$\exp(-16\xb[q+\xf])>0$, and $[1-\exp\left(-32\xf\xb\right)]^{-1}>1$.

\section{Structure of resonance states}
\label{App:StructureOfResonanceStates}

In this appendix we discuss the asymptotic structure of resonance states in
Floquet scattering systems.
We further discuss its relation to the phenomenon of covering and uncovering of
resonances by crossings of the continuous spectrum.

We start by discussing the asymptotic structure of Floquet solutions for
$\theta=0$.
To this end we take, Eq.~\eqref{eq:FloquetSolutionExpandedCS}, insert it into
the Schr{\"o}dinger equation~\eqref{eq:SchroedingerEquation}, and consider the
asymptotic regions $q\gg1$, where the potential vanishes.
This gives
\begin{align}
 \label{eq:SchroedingerFloquetAsymptotic}
 (\epsilon+\mu\hbar\omega) \phi_{\epsilon,\mu}(q) =
-\frac{\hbar^2}{2}\frac{\partial^2}{\partial q^2}\phi_{\epsilon,\mu}(q).
\end{align}
This shows that each component $\phi_{\epsilon,\mu}(q)$ is associated with an
asymptotic energy $\epsilon+\mu\hbar\omega$.
Its asymptotic form is a linear combination of solutions
\begin{align}
 \label{eq:ResonanceStateFloquetAsymptotic}
 \phi_{\epsilon,\mu}^{\pm}(q)
 \simeq
 \exp{\left(i p^{\pm}(\epsilon+\mu\hbar\omega) q/\hbar  \right)},
\end{align}
with momenta
\begin{align}
 \nonumber
 p^{\pm}(\epsilon+\mu\hbar\omega) = \pm\sqrt{|2(\epsilon+\mu\hbar\omega)|}
\exp{\left(i\frac{\Arg{(\epsilon+\mu\hbar\omega)}}{2}\right)}.
\end{align}
Here, we fix the branch cut of the energy root along the positive real axis,
i.e., the complex argument function $\Arg{(\cdot)}$ maps to values on the
interval $(0,2\pi]$.

Associated with scattering solution which have quasi-energies within the first
Floquet zone along the real axis $\epsilon\in[0,\hbar\omega)$ we introduce the
following classification:\
(i) Energies with $\mu\geq0$ correspond to open propagating channels, with
$p^{-}$ an outgoing and $p^{+}$ an incoming solution.
(ii) Solutions with $\mu<0$ correspond to closed channels, with $p^{+}$ an
exponentially decreasing evanescent and $p^{-}$ an exponentially increasing
solution.
In what follows we keep the terminology, even if $\epsilon$ is in the lower
half of the complex plane.

A general scattering solution of a Floquet system consists of both incoming and
outgoing solutions in the open channels and possibly evanescent components in
the closed channels.
If a solution is found which is purely outgoing in the open channels, we speak
of a resonance solution, which admits the following asymptotic form
\begin{align}
 \label{eq:ResonanceStateFloquetAsymptotic}
 \phi_{\epsilon,\mu}(q)
 \simeq
 \begin{cases}
 \exp{\left(i p^{-}(\epsilon+\mu\hbar\omega) q/\hbar  \right)} \quad
&\mu\ge0,\\
  \exp{\left(i p^{+}(\epsilon+\mu\hbar\omega) q/\hbar\right)}
\quad &\mu<0.
 \end{cases}
\end{align}
A resonance states does generally not exist for real quasi-energies, but
exhibits $\epsilon$ in the lower half of the complex plane.
For this reason the outgoing solutions in the open channels diverge at infinity
such that a resonance state is not square-integrable and can thus not be
computed with a simple basis state expansion of the Floquet solution,
Eq.~\eqref{eq:FloquetSolutionExpandedCS}.

\begin{figure}[tb!]
  \begin{center}
    \includegraphics[]{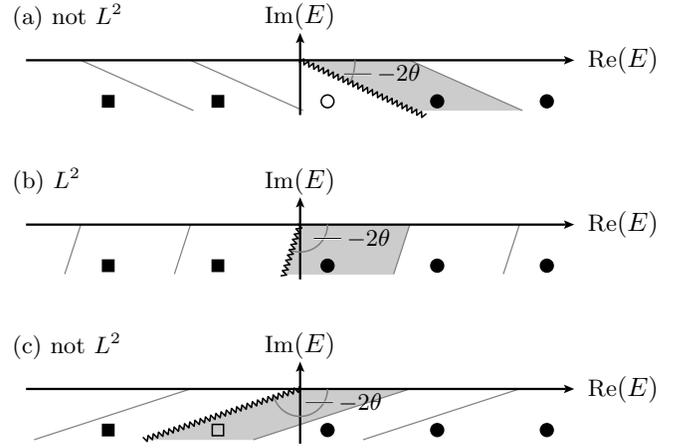}
     \caption{(color online) Square-integrability of a resonance state in a
Floquet system for increasing scaling angle $\theta$.
The asymptotic energies of its components are shown by squares and circles.
Circles denote open channels with outgoing solutions on the second Riemann
sheet.
Squares denote closed channels with localized solutions on the first Riemann
sheet.
Full symbols denote square-integrable components.
Open symbols denote components which are not square-integrable.
The branch cut (zig-zag line), separates square-integrable components, from
components which are not square-integrable, both on the first and the second
Riemann sheet.
Gray lines mark the Floquet zone and the first zone is mark by a gray shaded
region.
}
     \label{fig:components_TPSE_schematic}
  \end{center}
\end{figure}
Upon complex scaling the asymptotic form of the components of the Floquet
resonance solution becomes
\begin{align}
 \label{eq:ResonanceStateFloquetAsymptoticCS}
 \phi_{\epsilon,\mu}^{\theta}(s)
 \simeq
 \begin{cases}
 \exp{\left(i p^{-}(\epsilon+\mu\hbar\omega) s \exp{(i\theta)/\hbar}
 \right)} \quad &\mu\ge0,\\
  \exp{\left(i p^{+}(\epsilon+\mu\hbar\omega) s \exp{(i\theta)}/\hbar\right)}
\quad &\mu<0.
 \end{cases}
\end{align}
In order to have a square-integrable solution,
Eq.~\eqref{eq:FloquetSolutionExpandedCS}, one requires every component to be
square-integrable.
For this to be fulfilled, we require two conditions:\
(i) The scaling angle $\theta$ should be large enough such that all open
channels become localized.
This is fulfilled if $\Arg(\epsilon)\in(-2\theta,0)$.
(ii) The scaling angle $\theta$ should be small enough such that all closed
channels remain localized.
This is fulfilled if $\Arg(\epsilon-\hbar\omega)\in(-\pi,-2\theta)$.

Conversely, if condition (i) is violated, then there are still open
channels which are not yet square-integrable.
If condition (ii) is violated, then there are closed channels which are no
longer square integrable.
In that we effectively have a branch cut along the line
$\exp(-i2\theta)\mathbb{R}_{+}$, which separates square integrable solutions
from diverging solutions, both on the first and the second Riemann sheet.
See Fig.~\ref{fig:components_TPSE_schematic} for a sketch.

Note that conditions (i) and (ii) are equivalent to saying that a resonance
state has square-integrable components, if the quasi-energy $\epsilon$
associated with the lowest open channel is in the first Floquet zone,
\eqref{eq:FloquetZone}.
This explains why upon increasing the scaling angle a resonance in a Floquet
system can both be uncovered and later be covered again.

\section{Numerical methods}
\label{App:NumericalImplementation}

In this section we describe the numerical computation of eigenvalues for
operators in Eq.~\eqref{eq:PropagatorScaled} and
Eq.~\eqref{eq:PropagatorAbsorption}.
The scheme is based on a finite element representation of the wave function and
follows the ideas of Refs.~\cite{ScrEla1993, Scr2010}.
The benefit of using a finite-element scheme is to achieve numerically
determined eigenvalues with high accuracy.
We describe the representation of the wave function in
Sec.~\ref{App:Wavefunction} and the representation of operators in
Sec.~\ref{App:Operators}.
Details specific to complex scaling, weak absorbing potentials, and projective
openings are discussed in Secs.~\ref{App:ComplexScaling},
\ref{App:AbsorbingPotential}, and \ref{App:ProjectiveOpening}.

\subsection{Representation of wave functions}
\label{App:Wavefunction}

We now introduce the finite element representation of wave-functions $\psi(x)$.
Here, $x$ may either refer to the coordinate $s$ of
Sec.~\ref{Sec:ComplexScaling} or $q$ of Sec.~\ref{Sec:AbsorbingPotential}.

We start by introducing a grid in coordinate space using grid points
\begin{align}
 \label{eq:grid}
 x_{0} < x_{1} < \dots < x_{N-1} \quad \text{with} \quad x_{l}\in\mathbb{R}.
\end{align}
We fix the extent of the grid as
\begin{align}
 \label{eq:gridsize}
 \xmax := x_{N-1}= -x_{0}.
\end{align}
We define the $l$th cell of the discretization as
\begin{align}
 \label{eq:cell}
 x\in[x_{l}, x_{l+1}].
\end{align}

\subsubsection{Basis functions in the $l$th cell}

We now introduce basis functions, which are non-zero only in the $l$th cell.
We start from Legendre polynomials on the interval $[-1,1]$, as defined by the
recursion
\begin{align}
  (n+1)\bar{P}_{n+1}(y) = (2n+1)y \bar{P}_{n}(y) - n \bar{P}_{n-1}(y).
\end{align}
with $\bar{P}_{0} = 1$ and $\bar{P}_{1} = y$.
The polynomial fulfill $P_{n}(1)=1$, a symmetry property
$P_{n}(y)=(-1)^{n}P_{n}(-x)$, and an orthogonality relation
\begin{align}
 \label{eq:OrthogonalityLegendre}
 \int_{-1}^{1}P_{n}(y)P_{m}(y)\text{d}y&= \frac{2}{2n+1}\delta_{m,n},
\end{align}
where $\delta_{m,n}$ is the Kronecker delta function.
From these polynomials we introduce a new set of polynomials
\begin{align}
  \label{eq:DefinitionChiBar}
  \bar{\chi}_{n}(y) :=
  \begin{cases}
    \frac{1}{2}\bar{P}_{0}(y) - \frac{1}{2}\bar{P}_{1}(y)
        &\text{if } n\!=\!0, \\
    \bar{P}_{n+1}(y) - \bar{P}_{\text{mod2}(n+1)}(y)
        &\text{if } n\!=\!1,\dots,\mmax\!-\!1,\\
    \frac{1}{2}\bar{P}_{0}(y) + \frac{1}{2}\bar{P}_{1}(y)
        &\text{if } n\!=\!\mmax, \\
   \end{cases}
\end{align}
which vanish at the edges of the interval $[-1,1]$.
The only exception being $\bar{\chi}_{0}(-1)=\bar{\chi}_{\mmax}(1)=1$.

We scale these polynomials to the $l$th cell using the linear transformation
\begin{align}
 y^{(l)}(x) := \frac{2x-(x_{l}+x_{l+1})}{x_{l+1}-x_{l}}
\end{align}
resulting in the scaled basis-functions on the $l$th cell
\begin{align}
 \chi_{n}^{(l)}(x):= \bar{\chi}_{n}(y^{(l)}(x)).
\end{align}
We emphasize that $\chi_{n}^{(l)}$ is defined to be zero outside of the $l$th
cell.
Note that for all cells $l=0,\dots,N-1$ we have:
\begin{subequations}
  \label{eq:ChiOnCellEdge}
  \begin{align}
  \chi_{n}^{(l)}(x_{l}) &= \begin{cases}
                          1 \quad &\text{if } n=0\\
                          0 \quad &\text{else},
                         \end{cases}\\
  \chi_{n}^{(l)}(x_{l+1}) &= \begin{cases}
                          1 &\quad \text{if } n=\mmax\\
                          0 &\quad \text{else}.
                         \end{cases}
  \end{align}
\end{subequations}

\subsubsection{Wavefunction}

We can now write the wave function along the grid as
\begin{align}
 \label{eq:TrialWavefunction}
 \psi(x) = \sum_{l=0}^{N-1} \sum_{n=0}^{\mmax} c_{n}^{(l)}\chi_{n}^{(l)}(x).
\end{align}
In order to ensure the continuity of this wave-function, we exploit property
\eqref{eq:ChiOnCellEdge}, and impose an additional constraint
\begin{align}
  \label{eq:DependentCoefficients}
  c_{\mmax}^{(l)} = c_{0}^{(l+1)} \quad \forall l=0,\dots,N-2.
\end{align}

In order to handle dependent and independent coefficients, we arrange
the dependent coefficients into a vector
\begin{align}
 \begin{pmatrix}
  c_{0} \\
  c_{1}\\
  \vdots\\
  \vdots\\
  \vdots\\
  c_{\alphad}\\
  \vdots\\
  \vdots\\
  \vdots\\
  c_{\alphadmax} \\
 \end{pmatrix}
 =
 \begin{pmatrix}
  c_{0}^{(0)} \\
  \vdots\\
  c_{\mmax}^{(0)} \\
  c_{0}^{(1)} \\
  \vdots\\
  c_{\mmax}^{(1)} \\
  \vdots\\
  c_{0}^{(N-1)} \\
  \vdots\\
  c_{\mmax}^{(N-1)} \\
  \end{pmatrix}.
\end{align}
Here, the indexes referring to the $l$th cell and the $n$th basis function have
been mapped to a single index
\begin{align}
 \alphad(l,n) := (\mmax+1) l + n
\end{align}
with
\begin{align}
 \alphadmax = N(\mmax+1)-1.
\end{align}
The original indexes may be recovered as
\begin{align}
  n(\alphad) &= \alphad \mod (\mmax+1) \\
  l(\alphad) &= (\alphad - n(\alphad))/ (\mmax+1)
\end{align}

Similarly, we eliminate the dependent coefficients $c_{\mmax}^{(l)}$ for
$l=0,\dots,N-2$ via Eq.~\eqref{eq:DependentCoefficients} and arrange the
remaining independent coefficients into a vector
\begin{align}
 \begin{pmatrix}
  c_{0} \\
  c_{1}\\
  \vdots\\
  \vdots\\
  \vdots\\
  c_{\nui}\\
  \vdots\\
  \vdots\\
  \vdots\\
  c_{\nuimax} \\
 \end{pmatrix}
 =
 \begin{pmatrix}
  c_{0}^{(0)} \\
  \vdots\\
  c_{\mmax-1}^{(0)} \\
  c_{0}^{(1)} \\
  \vdots\\
  c_{\mmax-1}^{(1)} \\
  \vdots\\
  c_{0}^{(N-1)} \\
  \vdots\\
  c_{\mmax}^{(N-1)} \\
  \end{pmatrix}
\end{align}
Here, the indexes referring to the $l$th cell and the $n$th basis function have
been mapped to a single index
\begin{align}
 \nui(l,n) := \mmax l + n,
\end{align}
with
\begin{align}
  \nuimax = N\mmax+1.
\end{align}
Based on these indexes we can define the matrix which maps the vector of
independent coefficients to a vector of dependent coefficients
\begin{align}
  c_{\alphad} = \sum_{\nui=0}^{\nuimax} R_{\alphad,\nui}c_{\nui}.
\end{align}
These matrices have a very simple structure
\begin{align}
 \begin{pmatrix}
  1 &   0    & 0 &   &   &        &        \\
  0 & \ddots & 0 &   &   &        &        \\
  0 &    0   & 1 &   &   &        &        \\
    &        & 1 &      0 & 0 &        &        \\
    &        & 0 & \ddots & 0 &        &        \\
    &        & 0 &   0    & 1 &        &        \\
  \vdots &  \vdots      & \vdots  &  \vdots      & \vdots & \vdots  & \vdots
  \\
    &        &   &        & 1 &        &        \\
    &        &   &        &   & \ddots &        \\
    &        &   &        &   &        &  1     \\
    \end{pmatrix}
\end{align}

In terms of such matrices, we can define the independent basis functions
\begin{align}
 \label{eq:BasisFunctionIndependent}
 \chi_{\nui}(x) := \sum_{\alphad=0}^{\alphadmax}\chi_{\alphad}(x) R_{\alphad,
\nui},
\end{align}
and the wave function in terms of the independent basis functions is given by
\begin{align}
 \label{eq:TrialWavefunctionIndependent}
 \psi(x) = \sum_{\nui=0}^{\nuimax} c_{\nui} \chi_{\nui}(x).
\end{align}

\subsection{Representation of operators}
\label{App:Operators}

We deal with time-evolution equations of the form
\begin{align}
  \label{eq:AppTimeEvolution}
  \psi'(x') = \int U(x', x)\,\psi(x) \text{d}x,
\end{align}
where $U(x', x)$ denotes an operator $\braOpket{x'}{\qmap}{x}$.
We discretize such operators by expressing the wave functions as
\begin{align}
 \label{eq:AppPsiInitial}
 \psi(x)&=\sum_{\nui=0}^{\nuimax} c_{\nui}\chi_{\nui}(x),\\
 \label{eq:AppPsiFinal}
 \psi'(x')&=\sum_{\nui=0}^{\nuimax} d_{\nui}\chi_{\nui}(x'),
\end{align}
and testing against independent basis functions $\chi_{\nuibar}$.
This reduces Eq.~\eqref{eq:AppTimeEvolution} to
\begin{align}
  \label{eq:AppTimeEvolutionMatrix}
  \sum_{\nui=0}^{\nuimax} O_{\nuibar,\nui} d_{\nui} = \sum_{\nui=0}^{\nuimax}
U_{\nuibar,\nui} c_{\nui}.
\end{align}
Here, $U_{\nuibar,\nui}$ is the matrix representation of the operator $U(x',x)$
and $O_{\nuibar,\nui}$ is the overlap matrix.

The matrix representation of $U(x',x)$ is given by
\begin{align}
 \label{eq:MatrixIndependent}
 U_{\nuibar,\nui} := \int \text{d}x' \int \text{d}x \,\,
\chi_{\nuibar}(x')U(x',x)\chi_{\nui}(x)
\end{align}
This expression can be recast into the form
\begin{align}
 \label{eq:MatrixReduction}
 U_{\nuibar,\nui} := \sum_{\alphadbar=0}^{\alphadmax}
\sum_{\alphad=0}^{\alphadmax} R_{\alphadbar,\nuibar}
U_{\alphadbar,\alphad} R_{\alphad,\nui},
\end{align}
based on Eq.~\eqref{eq:BasisFunctionIndependent}.
Here, $U_{\alphadbar,\alphad}$ represents the operator $U(x',x)$ projected on
the basis set of dependent wave-functions
\begin{align}
 \label{eq:MatrixDependent}
 U_{\alphadbar,\alphad} := \int \text{d}x' \int \text{d}x \,\,
\chi_{\alphadbar}(x')U(x',x)\chi_{\alphad}(x).
\end{align}

The numerical scheme makes use of three types of matrices:
(i) the overlap matrix, (ii) kick-type time-evolution matrices, and (iii) a
matrix representation of the free propagator.

(i) The matrix representation of the overlap matrix is
\begin{align}
 \label{eq:AppOverlapMatrix}
 O_{\alphad,\alphadbar}
&=\int \text{d}x \, \chi_{\alphadbar}(x)\chi_{\alphad}(x)\\
&=\delta_{l(\alphadbar), l(\alphad)}
\int \text{d}x\,
\chi_{n(\alphad)}^{l(\alphad)}(x)\chi_{n(\alphadbar)}^{l(\alphadbar)}(x)\\
&=\delta_{l(\alphadbar), l(\alphad)} \frac{x_{l+1}-x_{l}}{2}
\int_{-1}^{1} \text{d}y\,
\bar{\chi}_{n(\alphad)}(y) \bar{\chi}_{n(\alphadbar)}(y)
\end{align}
The remaining integral can be computed from Eq.~\eqref{eq:DefinitionChiBar}
and Eq.~\eqref{eq:OrthogonalityLegendre}.
The expression shows that the basis states $\alphad,\alphadbar$ overlap, only
if they belong to the same cell $l(\alphad),l(\alphadbar)$.
Reducing the overlap matrix to independent coefficients according to
Eq.~\eqref{eq:MatrixReduction} results in a globally connected overlap matrix
$O$.
The overlap-matrix $O$ is real symmetric.
Its inverse $O^{-1}$ can be computed by diagonalizing $O$ and inverting its
eigenvalues.

(ii) The kick-type time-evolution operators arise as a multiplicative part in
Eqs.~\eqref{eq:PropagatorScaled} and \eqref{eq:PropagatorAbsorption}.
We summarize them in a unified form
\begin{align}
 \label{eq:AppKick}
 U^{V}(x,x') = \delta(x-x') \exp{\left(-\frac{i}{2\hbar} V(x) \right)},
\end{align}
where $V(x)$ denotes either the scaled potential or the kick potential
including absorption.
Its matrix representation is given by
\begin{align}
 \label{eq:AppKickMatrix}
 U_{\alphadbar,\alphad}^{V} =
 \delta_{l(\alphadbar), l(\alphad)}
 \int\!\!\text{d}x \,\chi_{n(\alphad)}^{l(\alphad)}(x)
\exp{\left(-\frac{i}{2\hbar} V(x) \right)}
\chi_{n(\alphadbar)}^{l(\alphadbar)}(x).
\end{align}
The remaining integral has to be evaluated numerically, as discussed below.
Similar to the overlap matrix, the matrix elements are non-zero only for matrix
elements of basis functions in the same cell.
The above matrix is reduced to a matrix $U^{V}$ on independent basis functions
according to Eq.~\eqref{eq:MatrixReduction}.
The corresponding coefficients can be propagated via $O^{-1}U^{V}$, according
to Eq.~\eqref{eq:AppTimeEvolutionMatrix}.

(iii) Finally, the matrix representation of propagators corresponding to the
free time evolution in between two kicks, is given by
\begin{widetext}
\begin{align}
 \label{eq:AppFreeMatrix}
 U_{\alphadbar,\alphad}^{T} =
 \frac{\exp{(i[\theta-\pi/4])}}{(2\pi\hbar)^{1/2}}
 \int\!\!\text{d}x'\int\!\!\text{d}x \,\chi_{n(\alphad)}^{l(\alphad)}(x')
\exp{\left(\frac{i}{\hbar}e^{i2\theta}\frac{(x-x')^2}{2}\right)}
\chi_{n(\alphadbar)}^{l(\alphadbar)}(x),
\end{align}
\end{widetext}
The remaining integrals have to be evaluated numerically, as discussed below.
The above matrix is reduced to a matrix $U^{T}$ on independent basis functions
according to Eq.~\eqref{eq:MatrixReduction}.
The corresponding coefficients can be propagated via $O^{-1}U^{T}$, according
to Eq.~\eqref{eq:AppTimeEvolutionMatrix}.

Finally, the matrix representation of the full propagator $U$, corresponding to
Eq.~\eqref{eq:PropagatorScaled} or Eq.~\eqref{eq:PropagatorAbsorption},
respectively, is given by
\begin{align}
 U = O^{-1}U^{V}O^{-1}U^{T}O^{-1}U^{V}.
\end{align}
Its eigenvalues are obtained from numerical diagonalization.

\subsection{Integration}

To ensure accurate matrix elements in Eq.~\eqref{eq:AppKickMatrix} and
Eq.~\eqref{eq:AppFreeMatrix}, we set the length of the cells such that the
integrand does not exhibit more than one oscillation within each cell.
We then perform the remaining numerical integration using Gaussian Integration
with 30-50 integration points, see Chapter 3.6 of Ref.~\cite{StoBul2002} for
details.

In practice we estimate the interval $[x,x+\Delta x]$ on which the integrands in
Eq.~\eqref{eq:AppKickMatrix} and Eq.~\eqref{eq:AppFreeMatrix} oscillate once by
considering their phase function in the limit of small $\hbar$.
This gives
\begin{align}
 \Delta x \simeq \frac{\heff}{V'(x)}\\
 \Delta x \simeq \frac{\heff}{|x-x'|}.
\end{align}
Both estimates show that the cell size scales with $\heff$.
Estimating the maximum of $V'(x)$ along the real axis (which is not entirely
correct for the scaled potential) from its main part $V_{\kappa}(q)$, we obtain
\begin{align}
   \max\{V'(x)\} \simeq \frac{\kappa}{4}\exp{\left(-\frac{1}{2}\right)}.
\end{align}
On the other hand, we estimate the maximal relevant length $|x-x'|$ in the
integration of Eq.~\eqref{eq:AppFreeMatrix} as
\begin{align}
 \max\{|x-x'|\} \simeq
 \begin{cases}
  \sqrt{\frac{30\hbar\log(10)}{\sin(2\theta)}}, \\
  2\qabs,
 \end{cases}
\end{align}
in the case of complex scaling and absorbing potentials, respectively.
The first estimate is obtained by computing the separation $|x-x'|$ for which
the non-oscillatory part of the kernel in Eq.~\eqref{eq:AppFreeMatrix} drops
below machine precision.
The second estimate is obtained by assuming that significant contribution
in Eq.~\eqref{eq:AppFreeMatrix} arise mainly between points in the
absorption-free region $q\in[-\qabs,\qabs]$.
Usually we have $\max\{|x-x'|\}>\max\{V'(x)\}$ and thus set the number of
cells in an equidistant grid as
\begin{align}
 \label{eq:NGrid}
 N  \approx \frac{1}{h} \times
 \begin{cases}
  \sqrt{\frac{30\hbar\log(10)}{\sin(2\theta)}}, \\
  2\qabs,
 \end{cases}
\end{align}
in the case of complex scaling and absorbing potentials, respectively.

\subsection{Testing}

In order to confirm the correctness of the finite-element representation of
wave functions, we first applied it to several time-independent problems.
These include the box potential, the harmonic oscillator, and the Bain
potential, discussed in Ref.~\cite{ScrEla1993}.
In all cases we could confirm the relevant eigenvalues up to machine precision.
This test gives confidence to the correctness of the overlap matrix as well as
the numerical integration routines.

Subsequently, we focus on the matrices $U_{V}$, deduced from
Eq.~\eqref{eq:AppKickMatrix}.
Here, we test the numerical integrations in Eq.~\eqref{eq:AppKickMatrix} by
comparing to analytical results obtained for simple basis functions.
We further tested the matrix $U^{V}$ by comparing that
(i) reducing a known wave functions $\psi(x)$ to coefficients,
(ii) propagating the coefficients by $O^{-1}U^{V}$, and
(iii) reconstructing the propagated wave-function $\psi'(x')$ from the
propagated coefficients,
agrees well with the expression $\psi(x)\exp(-iV(x)/(2\hbar))$ within the
maximum norm.

In a similar way we test the matrix $U_{T}$, deduced from
Eq.~\eqref{eq:AppFreeMatrix}.
We test the numerical integrations in Eq.~\eqref{eq:AppFreeMatrix} by
comparing to analytical results obtained for simple basis functions.
We further tested the matrix $U^{T}$ by comparing that
(i) reducing a Gaussian wave functions $\psi_{g}(x)$ to coefficients,
(ii) propagating the coefficients by $O^{-1}U^{T}$, and
(iii) reconstructing the propagated wave-function $\psi'_{g}(x',t=1)$ from
propagated coefficients,
agrees well within the maximum norm with the closed form analytical expressions
\begin{widetext}
\begin{align}
 \psi_{g}(x,t) &=
\frac{1}{(2\pi)^{1/4}}\frac{\sigma_{0}^{1/2}}{\sigma_{t}}\exp{\left(-\frac{
(x-x_{t})^2}{4\sigma_{t}^2}  +i\frac{p_{0}}{\hbar}(x-x_{t})
+i\frac{p_{0}x_{t}}{2\hbar} \right) },
\end{align}
\end{widetext}
with
\begin{align}
x_{t} &= x_{0} + p_{0} t \exp(-i2\theta), \\
\sigma_{t} &= \sqrt{\sigma_{0}^2 + \frac{i\hbar}{2}t\exp(-i2\theta)}
\end{align}
obtained when propagating a Gaussian with the complex-scaled propagator,
Eq.~\eqref{eq:FreePropagatorScaled}.

\subsection{Complex scaling}
\label{App:ComplexScaling}

The numerical computations for complex scaling in this paper use a grid with
$\xmax=4.0$.
The order of the basis functions in each cell was $\mmax=10$.
The cell width is equidistant.
The number of cells was determined as $N=840$ for the computation with
$\theta=0.05$ and $N=665$ for $\theta=0.08$.

All parameters used in this computation were aimed at high accuracy.
In particular, further increasing $\xmax$, $\mmax$ or $N$ does not change the
spectra.
On the contrary, if resonance spectra are required up to few significant digits
only, the same resonance eigenvalues can be computed for much smaller sets of
basis functions $\mmax$ and slightly smaller grids $\xmax$.

We cross examined the numerical result by computing the spectrum of the complex
scaled quantum map via a quantum maps code, i.e., putting the system on a
torus, setting up the operator via split-operator FFT, and computing the
spectrum numerically.
This gives the same resonance spectrum within several significant digits.

Note that for complex scaling the operator in Eq.~\eqref{eq:AppKick} can be
amplifying for some positions and damping for others.
In the semiclassical limit $\hbar\to0$ this effects become exponentially
amplified, leading to matrix elements in Eq.~\eqref{eq:AppKickMatrix} which span
many orders of magnitude.
Eventually, this can make the numerical scheme unstable.
If this happens the scaling angle $\theta$ must be lowered.

\subsection{Projective opening}
\label{App:ProjectiveOpening}

The numerical computations with projective openings use basis functions in each
cell up to order $\mmax=5$.
The cell width is equidistant.
The number of cells was determined as $N=1200$ for the computation with
$\xmax=2.0$ and $N=1500$ for $\xmax=2.5$.
In both cases we cannot identify resonance eigenvalues which agree with the
results of complex scaling within several significant digits.

The convergence of the spectra was checked, i.e., doubling the number of
cells $N$ or the order $\mmax$ or both gives the same spectra.
Note that treating the projective opening with a quantum map code, i.e.,
imposing periodic boundary conditions in both position and momentum gives
additional spurious modes.

\subsection{Absorbing potential}
\label{App:AbsorbingPotential}

The numerical computations for absorbing potentials in this paper use basis
functions in each cell up to order $\mmax=3$.
The cell width is equidistant.
The number of cells was determined as $N=1500$ for the computation with
$\xmax=2.5$ and $N=3000$ for $\xmax=5$.
Spectra obtained in the case of weak absorption $\xmax=5.0$ give the same
resonance eigenvalues as complex scaling within several significant digits.
On the contrary the case of strong absorbing potentials $\xmax=2.5$ does not
agree with resonance eigenvalues obtained from complex scaling.
We further checked the convergence of the spectra, i.e., doubling the number of
cells $N$ or the order $\mmax$ or both gives the same spectra.
Note that treating the absorbing potential case with a quantum map code, i.e.,
imposing periodic boundary conditions in both position and momentum gives
additional spurious modes.

\subsubsection{Smooth monomials}
\label{App:SmoothMonomials}

We define the smooth monomials of Sec.~\ref{Sec:AbsorbingPotential} as
\begin{align}
 M_{n}^{\width}(x) := \width^{n} F_{n}(x/\width),
\end{align}
where $\width$ is a smoothness parameter and the functions $F_{n}(\cdot)$ up to
order $n=3$ are given by
\begin{align}
 F_{-1}(x)     &= \frac{\exp{(-x^2)}}{\pi^{1/2}} \\
 F_{0}(x) &= \frac{1}{2}\left(1+\erf(x)\right) \\
 F_{1}(x) &= x F_{0}(x) + \frac{1}{2} F_{-1}(x)\\
 F_{2}(x) &= \left(x^2 + \frac{1}{2}\right) F_{0}(x) + \frac{x}{2} F_{-1}(x)\\
 F_{3}(x) &= \left(x^3 + \frac{3x}{2}\right) F_{0}(x) +
             \left(\frac{x^2}{2} + \frac{1}{2}\right) F_{-1}(x).
\end{align}

\end{appendix}


\begin{thebibliography}{10}
\newcommand{\enquote}[1]{``#1''}
\providecommand{\url}[1]{\texttt{#1}}
\providecommand{\urlprefix}{URL }
\providecommand{\eprint}[2][]{\url{#2}}

\bibitem{GasRic1989a}
P.~Gaspard and S.~A. Rice, \emph{Scattering from a classically chaotic
  repellor}, J.~Chem.~Phys. \textbf{90}, 2225 (1989).

\bibitem{GasRic1989b}
P.~Gaspard and S.~A. Rice, \emph{Semiclassical quantization of the scattering
  from a classically chaotic repellor}, J.~Chem.~Phys. \textbf{90}, 2242
  (1989).

\bibitem{GasRic1989c}
P.~Gaspard and S.~A. Rice, \emph{Exact quantization of the scattering from a
  classically chaotic repellor}, J.~Chem.~Phys. \textbf{90}, 2255 (1989).

\bibitem{BluSmi1989}
R.~Bl\"umel and U.~Smilansky, \emph{A simple model for chaotic scattering:
  {II}. {Quantum} mechanical theory}, Physica~D \textbf{36}, 111 (1989).

\bibitem{BluSmi1990}
R.~Bl{\"u}mel and U.~Smilansky, \emph{Random-matrix description of chaotic
  scattering: Semiclassical approach}, Phys.~Rev.~Lett. \textbf{64}, 241
  (1990).

\bibitem{Jun1990}
C.~Jung, \emph{Fractal properties in the semiclassical scattering cross section
  of a classically chaotic system}, J.~Phys.~A \textbf{23}, 1217 (1990).

\bibitem{JunPot1990}
C.~Jung and S.~Pott, \emph{Semiclassical cross section for a classically
  chaotic scattering system}, J.~Phys.~A \textbf{23}, 3729 (1990).

\bibitem{JunTel1991}
C.~Jung and T.~T\'el, \emph{Dimension and escape rate of chaotic scattering
  from classical and semiclassical cross section data}, J.~Phys.~A \textbf{24},
  2793 (1991).

\bibitem{CviEck1989}
P.~Cvitanovi{\'c} and B.~Eckhardt, \emph{Periodic-orbit quantization of chaotic
  systems}, Phys.~Rev.~Lett. \textbf{63}, 823 (1989).

\bibitem{TanRicRos2000}
G.~Tanner, K.~Richter, and J.-M. Rost, \emph{The theory of two-electron atoms:
  between ground state and complete fragmentation}, Rev.~Mod.~Phys.
  \textbf{72}, 497 (2000).

\bibitem{Sjo1990}
J.~Sj\"ostrand, \emph{Geometric bounds on the density of resonances for
  semiclassical problems}, Duke Math.~J. \textbf{60}, 1 (1990).

\bibitem{Lin2002}
K.~K. Lin, \emph{Numerical study of quantum resonances in chaotic scattering},
  J.~Comput.~Phys. \textbf{176}, 295 (2002).

\bibitem{LuSriZwo2003}
W.~T. Lu, S.~Sridhar, and M.~Zworski, \emph{Fractal weyl laws for chaotic open
  systems}, Phys.~Rev.~Lett. \textbf{91}, 154101 (2003).

\bibitem{SchTwo2004}
H.~Schomerus and J.~Tworzyd{\l}o, \emph{Quantum-to-classical crossover of
  quasibound states in open quantum systems}, Phys.~Rev.~Lett. \textbf{93},
  154102 (2004).

\bibitem{Non2011}
S.~Nonnenmacher, \emph{Spectral problems in open quantum chaos}, Nonlinearity
  \textbf{24}, R123 (2011).

\bibitem{Nov2013}
M.~Novaes, \emph{Resonances in open quantum maps}, J.~Phys.~A \textbf{46},
  143001 (2013).

\bibitem{KoeMicBaeKet2013}
M.~J. K\"orber, M.~Michler, A.~B\"acker, and R.~Ketzmerick, \emph{Hierarchical
  fractal {W}eyl laws for chaotic resonance states in open mixed systems},
  Phys.~Rev.~Lett. \textbf{111}, 114102 (2013).

\bibitem{KeaNovPraSie2006}
J.~P. Keating, M.~Novaes, S.~D. Prado, and M.~Sieber, \emph{Semiclassical
  structure of chaotic resonance eigenfunctions}, Phys.~Rev.~Lett. \textbf{97},
  150406 (2006).

\bibitem{KoeBaeKet2015}
M.~J. K\"orber, A.~B\"acker, and R.~Ketzmerick, \emph{Localization of chaotic
  resonance states due to a partial transport barrier}, Phys.~Rev.~Lett.
  \textbf{115}, 254101 (2015).

\bibitem{NonSjoZwo2011}
S.~Nonnenmacher, J.~Sj{\"o}strand, and M.~Zworski, \emph{From open quantum
  systems to open quantum maps}, Commun.~Math.~Phys. \textbf{304}, 1 (2011).

\bibitem{CasMasShe1997}
G.~Casati, G.~Maspero, and D.~L. Shepelyansky, \emph{Relaxation process in a
  regime of quantum chaos}, Phys.~Rev.~E \textbf{56}, R6233 (1997).

\bibitem{CasGuaMas2000}
G.~Casati, I.~Guarneri, and G.~Maspero, \emph{Fractal survival probability
  fluctuations}, Phys.~Rev.~Lett. \textbf{84}, 63 (2000).

\bibitem{NonZwo2005}
S.~Nonnenmacher and M.~Zworski, \emph{Fractal weyl laws in discrete models of
  chaotic scattering}, J.~Phys.~A \textbf{38}, 10683 (2005).

\bibitem{NonZwo2007}
S.~Nonnenmacher and M.~Zworski, \emph{Distribution of resonances for open
  quantum maps}, Commun.~Math.~Phys. \textbf{269}, 311 (2007).

\bibitem{NonRub2007}
S.~Nonnenmacher and M.~Rubin, \emph{Resonant eigenstates for a quantized
  chaotic system}, Nonlinearity \textbf{20}, 1387 (2007).

\bibitem{KeaNovSch2008}
J.~P. Keating, M.~Novaes, and H.~Schomerus, \emph{Model for chaotic dielectric
  microresonators}, Phys.~Rev.~A \textbf{77}, 013834 (2008).

\bibitem{KeaNonNovSie2008}
J.~P. Keating, S.~Nonnenmacher, M.~Novaes, and M.~Sieber, \emph{On the
  resonance eigenstates of an open quantum baker map}, Nonlinearity
  \textbf{21}, 2591 (2008).

\bibitem{NovPedWisCarKea2009}
M.~Novaes, J.~M. Pedrosa, D.~Wisniacki, G.~G. Carlo, and J.~P. Keating,
  \emph{Quantum chaotic resonances from short periodic orbits}, Phys.~Rev.~E
  \textbf{80}, 035202 (2009).

\bibitem{KopSch2010}
M.~Kopp and H.~Schomerus, \emph{Fractal {Weyl} laws for quantum decay in
  dynamical systems with a mixed phase space}, Phys.~Rev.~E \textbf{81}, 026208
  (2010).

\bibitem{IshAkaShuSch2012}
A.~Ishii, A.~Akaishi, A.~Shudo, and H.~Schomerus, \emph{Weyl law for open
  systems with sharply divided mixed phase space}, Phys.~Rev.~E \textbf{85},
  046203 (2012).

\bibitem{LipRyuLeeKim2012}
D.~Lippolis, J.-W. Ryu, S.-Y. Lee, and S.~W. Kim, \emph{On-manifold
  localization in open quantum maps}, Phys.~Rev.~E \textbf{86}, 066213 (2012).

\bibitem{SchAlt2015}
M.~Sch\"onwetter and E.~G. Altmann, \emph{Quantum signatures of classical
  multifractal measures}, Phys.~Rev.~E \textbf{91}, 012919 (2015).

\bibitem{MerKulLoeBaeKet2016}
N.~Mertig, J.~Kullig, C.~L{\"o}bner, A.~B{\"a}cker, and R.~Ketzmerick,
  \emph{Perturbation-free prediction of resonance-assisted tunneling in mixed
  regular-chaotic systems}, Phys.~Rev.~E \textbf{94}, 062220 (2016).

\bibitem{FriBaeKetMer2017}
F.~Fritzsch, A.~B{\"a}cker, R.~Ketzmerick, and N.~Mertig, \emph{Complex-path
  prediction of resonance-assisted tunneling in mixed systems}, Phys.~Rev.~E
  \textbf{95}, 020202(R) (2017).

\bibitem{NonSjoZwo2011:p}
S.~Nonnenmacher, J.~Sj{\"o}strand, and M.~Zworski, \emph{Fractal weyl law for
  open quantum chaotic maps}, arXiv:1105.3128 [math nlin]  (2011).

\bibitem{Sch2013b}
H.~Schomerus, \emph{From scattering theory to complex wave dynamics in
  non-hermitian {$\mathcal{PT}$}-symmetric resonators}, Phil.~Trans.~R.~Soc.~A
  \textbf{371}, 20120194 (2013).

\bibitem{Fes1958}
H.~Feshbach, \emph{Unified theory of nuclear reactions}, Ann.~Phys. \textbf{5},
  357 (1958).

\bibitem{Fes1962}
H.~Feshbach, \emph{A unified theory of nuclear reactions. ii}, Ann.~Phys.
  \textbf{19}, 287 (1962).

\bibitem{Rot2009}
I.~Rotter, \emph{A non-{H}ermitian {H}amilton operator and the physics of open
  quantum systems}, J.~Phys.~A \textbf{42}, 153001 (2009).

\bibitem{RisMey1993}
U.~V. Riss and H.-D. Meyer, \emph{Calculation of resonance energies and widths
  using the complex absorbing potential method}, J.~Phys.~B \textbf{26}, 4503
  (1993).

\bibitem{VatWirRos1994}
G.~Vattay, A.~Wirzba, and P.~E. Rosenqvist, \emph{Periodic orbit theory of
  diffraction}, Phys.~Rev.~Lett. \textbf{73}, 2304 (1994).

\bibitem{BreWirRotStaBur2010}
I.~B{\v{r}}ezinov{\'a}, L.~Wirtz, S.~Rotter, C.~Stampfer, and
  J.~Burgd{\"o}rfer, \emph{Transport through open quantum dots: Making
  semiclassics quantitative}, Phys.~Rev.~B \textbf{81}, 125308 (2010).

\bibitem{LacBreBurLib2013}
F.~Lackner, I.~B{\v{r}}ezinov{\'a}, J.~Burgd{\"o}rfer, and F.~Libisch,
  \emph{Semiclassical wave functions for open quantum billiards}, Phys.~Rev.~E
  \textbf{88}, 022916 (2013).

\bibitem{AguCom1971}
J.~Aguilar and J.~M. Combes, \emph{A {C}lass of {A}nalytic {P}erturbations for
  {O}ne-body {S}chr{\"o}dinger {H}amiltonians}, Commun. Math. Phys.
  \textbf{22}, 269 (1971).

\bibitem{BalCom1971}
E.~Balslev and J.~M. Combes, \emph{{S}pectral {P}roperties of {M}any-body
  {S}chr{\"o}dinger {O}perators with {D}ilatation-analytic {I}nteractions},
  Commun. Math. Phys. \textbf{22}, 280 (1971).

\bibitem{ReeSim1978}
M.~Reed and B.~Simon, \emph{Methods of Modern Mathematical Physic IV: Analysis
  of Operators}, Academic Press (1978).

\bibitem{Rei1982b}
W.~P. Reinhardt, \emph{Complex coordinates in the theory of atomic and
  molecular structure and dynamics}, Ann.~Rev.~Phys.~Chem. \textbf{33}, 223
  (1982).

\bibitem{Moi1998}
N.~Moiseyev, \emph{Quantum theory of resonances: calculating energies, widths
  and cross-sections by complex scaling}, Phys.~Rep. \textbf{302}, 212 (1998),
  00634.

\bibitem{ChuRei1977}
S.-I. Chu and W.~P. Reinhardt, \emph{Intense field multiphoton ionization via
  complex dressed states: Application to the h atom}, Phys.~Rev.~Lett.
  \textbf{39}, 1195 (1977).

\bibitem{MaqChuRei1983}
A.~Maquet, S.-I. Chu, and W.~P. Reinhardt, \emph{{Stark} ionization in dc and
  ac fields: An $l^2$ complex-coordinate approach}, Phys.~Rev.~A \textbf{27},
  2946 (1983).

\bibitem{MoiKor1990}
N.~Moiseyev and H.~J. Korsch, \emph{Metastable quasienergy positions and widths
  for time-periodic {H}amiltonians by the complex-coordinate method},
  Phys.~Rev.~A \textbf{41}, 498 (1990).

\bibitem{BenMoiLefKos1991}
N.~Ben-Tal, N.~Moiseyev, C.~Leforestier, and R.~Kosloff, \emph{Positions,
  lifetimes, and partial widths of metastable quasienergy states by solving the
  time-dependent complex-scaled schr{\"o}dinger equation}, J.~Chem.~Phys.
  \textbf{94}, 7311 (1991).

\bibitem{BenMoiKosCer1993}
N.~Ben-Tal, N.~Moiseyev, R.~Kosloff, and C.~Cerjan, \emph{Harmonic generation
  in ionizing systems by the time-dependent complex coordinate floquet method},
  J. Phys. B: At. Mol. Opt. Phys. \textbf{26}, 1445 (1993).

\bibitem{MacMeiPer1984}
R.~S. MacKay, J.~D. Meiss, and I.~C. Percival, \emph{Stochasticity and
  transport in {H}amiltonian systems}, Phys.~Rev.~Lett. \textbf{52}, 697
  (1984).

\bibitem{MacMeiPer1984b}
R.~S. Mackay, J.~D. Meiss, and I.~C. Percival, \emph{Transport in {H}amiltonian
  systems}, Physica~D \textbf{13}, 55 (1984).

\bibitem{Rom1990}
V.~Rom-Kedar, \emph{Transport rates of a class of two-dimensional maps and
  flows}, Physica~D \textbf{43}, 229 (1990).

\bibitem{Wig1992}
S.~Wiggins, \emph{Chaotic Transport in Dynamical Systems}, volume~2 of
  \emph{Interdisciplinary Applied Mathematics}, Springer New York (1992).

\bibitem{Rom1994}
V.~Rom-Kedar, \emph{Homoclinic tangles-classification and applications},
  Nonlinearity \textbf{7}, 441 (1994).

\bibitem{MitHanTigFloDel2004}
K.~A. Mitchell, J.~P. Handley, B.~Tighe, A.~Flower, and J.~B. Delos,
  \emph{Chaos-induced pulse trains in the ionization of hydrogen},
  Phys.~Rev.~Lett. \textbf{92}, 073001 (2004).

\bibitem{MitHanTigFloDel2004b}
K.~A. Mitchell, J.~P. Handley, B.~Tighe, A.~Flower, and J.~B. Delos,
  \emph{Analysis of chaos-induced pulse trains in the ionization of hydrogen},
  Phys.~Rev.~A \textbf{70}, 043407 (2004).

\bibitem{BurMitWykYeDun2011}
K.~Burke, K.~A. Mitchell, B.~Wyker, S.~Ye, and F.~B. Dunning,
  \emph{Demonstration of turnstiles as a chaotic ionization mechanism in
  rydberg atoms}, Phys.~Rev.~Lett. \textbf{107}, 113002 (2011).

\bibitem{DavGra1986}
M.~J. Davis and S.~K. Gray, \emph{Unimolecular reactions and phase space
  bottlenecks}, J.~Chem.~Phys. \textbf{84}, 5389 (1986).

\bibitem{Eck1987}
B.~Eckhardt, \emph{Fractal properties of scattering singularities}, J.~Phys.~A
  \textbf{20}, 5971 (1987).

\bibitem{BerBalTabVor1979}
M.~V. Berry, N.~L. Balazs, M.~Tabor, and A.~Voros, \emph{Quantum maps},
  Ann.~Phys.~(N.Y.) \textbf{122}, 26 (1979).

\bibitem{CasChiIzrFor1979}
G.~Casati, B.~Chirikov, F.~Izrailev, and J.~Ford, \emph{Stochastic behavior of
  a quantum pendulum under a periodic perturbation}, in G.~Casati and J.~Ford
  (editors) \enquote{Stochastic Behavior in Classical and Quantum Hamiltonian
  Systems}, volume~93 of \emph{Lect.~Notes Phys.}, 334, Springer Berlin /
  Heidelberg (1979).

\bibitem{Jen1992}
J.~H. Jensen, \emph{Quantum corrections for chaotic scattering}, Phys.~Rev.~A
  \textbf{45}, 8530 (1992).

\bibitem{Jen1995}
J.~H. Jensen, \emph{Accuracy of the semiclassical approximation for chaotic
  scattering}, Phys.~Rev.~E \textbf{51}, 1576 (1995).

\bibitem{KriFisOttAnt2011}
Y.~Krivolapov, S.~Fishman, E.~Ott, and T.~M. Antonsen, \emph{Quantum chaos of a
  mixed open system of kicked cold atoms}, Phys.~Rev.~E \textbf{83}, 016204
  (2011).

\bibitem{OniShuIkeTak2001}
T.~Onishi, A.~Shudo, K.~S. Ikeda, and K.~Takahashi, \emph{Tunneling mechanism
  due to chaos in a complex phase space}, Phys.~Rev.~E \textbf{64}, 025201
  (2001), 00054.

\bibitem{OniShuIkeTak2003}
T.~Onishi, A.~Shudo, K.~S. Ikeda, and K.~Takahashi, \emph{Semiclassical study
  on tunneling processes via complex-domain chaos}, Phys.~Rev.~E \textbf{68},
  056211 (2003).

\bibitem{Chi1979}
B.~V. {Chirikov}, \emph{{A universal instability of many-dimensional oscillator
  systems}}, Phys.~Rep. \textbf{52}, 263 (1979).

\bibitem{Sma1962}
S.~Smale, \emph{Dynamical systems and the topological conjugacy problem for
  diffeomorphisms}, Proc. Int. Congress Math. (Stockholm, 1962), Inst.
  Mittag-Leffler, Djursholm 490 (1962).

\bibitem{Wig2003}
S.~Wiggins, \emph{Introduction to Applied Nonlinear Dynamical Systems and
  Chaos}, number~2 in {Texts in Applied Mathematics}, {Springer New York}
  (2003).

\bibitem{Flo1883}
G.~Floquet, \emph{Sur les {\'e}quations diff{\'e}rentielles lin{\'e}aires {\'a}
  coefficients p{\'e}riodiques}, Annales scientifiques de l'{\'E}cole {N}ormale
  {S}up{\'e}rieure, S{\'e}rie 2 \textbf{12}, 47 (1883).

\bibitem{Sam1973}
H.~Sambe, \emph{Steady states and quasienergies of a quantum-mechanical system
  in an oscillating field}, Phys.~Rev.~A \textbf{7}, 2203 (1973), 00539.

\bibitem{Yaj1982}
K.~Yajima, \emph{Resonances for the {AC}-{Stark} effect}, Commun. Math. Phys.
  \textbf{87}, 331 (1982).

\bibitem{GraYaj1983}
S.~Graffi and K.~Yajima, \emph{Exterior complex scaling and the {AC}-{Stark}
  effect in a coulomb field}, Commun. Math. Phys. \textbf{89}, 277 (1983).

\bibitem{How1983}
J.~S. Howland, \emph{Complex scaling of ac {Stark} {H}amiltonians},
  J.~Math.~Phys. \textbf{24}, 1240 (1983).

\bibitem{Sim1973}
B.~Simon, \emph{Resonances in n-body quantum systems with dilatation analytic
  potentials and the foundations of time-dependent perturbation theory}, Annals
  of Mathematics \textbf{97}, 247 (1973).

\bibitem{Sim1978}
B.~Simon, \emph{Resonances and complex scaling: A rigorous overview},
  International Journal of Quantum Chemistry \textbf{14}, 529 (1978).

\bibitem{ScrEla1993}
A.~Scrinzi and N.~Elander, \emph{A finite element implementation of exterior
  complex scaling for the accurate determination of resonance energies},
  J.~Chem.~Phys. \textbf{98}, 3866 (1993).

\bibitem{Scr2010}
A.~Scrinzi, \emph{Infinite-range exterior complex scaling as a perfect absorber
  in time-dependent problems}, Phys.~Rev.~A \textbf{81}, 053845 (2010).

\bibitem{HanBer1980}
J.~H. Hannay and M.~V. Berry, \emph{Quantization of linear maps on a torus ---
  {F}resnel diffraction by a periodic grating}, Physica~D \textbf{1}, 267
  (1980).

\bibitem{DegGra2003}
M.~Degli~Esposti and S.~Graffi (editors) \emph{The Mathematical Aspects of
  Quantum Maps}, volume 618 of \emph{Lect.~Notes Phys.}, Springer-Verlag,
  Berlin (2003).

\bibitem{RamPraBorFar2009}
J.~A. Ramilowski, S.~D. Prado, F.~Borondo, and D.~Farrelly, \emph{Fractal weyl
  law behavior in an open hamiltonian system}, Phys.~Rev.~E \textbf{80}, 055201
  (2009).

\bibitem{Bog1992}
E.~B. Bogomolny, \emph{{Semiclassical quantization of multidimensional
  systems}}, Nonlinearity \textbf{5}, 805 (1992).

\bibitem{AltPorTel2013}
E.~G. Altmann, J.~S.~E. Portela, and T.~T\'el, \emph{Leaking chaotic systems},
  Rev.~Mod.~Phys. \textbf{85}, 869 (2013).

\bibitem{EckRue1985}
J.~P. Eckmann and D.~Ruelle, \emph{Ergodic theory of chaos and strange
  attractors}, Rev.~Mod.~Phys. \textbf{57}, 617 (1985).

\bibitem{PesMoi1994}
U.~Peskin and N.~Moiseyev, \emph{Time-independent scattering theory for
  time-periodic {H}amiltonians: Formulation and complex-scaling calculations of
  above-threshold-ionization spectra}, Phys.~Rev.~A \textbf{49}, 3712 (1994).

\bibitem{StoBul2002}
J.~Stoer and R.~Bulirsch, \emph{Introduction to {N}umerical {A}nalysis},
  volume~12, {S}pringer {N}ew {Y}ork, 3rd edition (2002).

\end{thebibliography}
\end{document}